\documentclass[aps,prb,twocolumn,floatfix,superscriptaddress,longbibliography]{revtex4-2}
\usepackage{lipsum}
\usepackage{bm}
\usepackage{subcaption}
\usepackage{colordvi}
\usepackage{color}
\usepackage{comment}
\usepackage{amssymb}
\usepackage{babel}
\usepackage[latin1]{inputenc}
\usepackage{amsbsy, amstext, amscd, amsxtra, amsopn, amsfonts, amsthm, amsmath}
\usepackage{hyperref}
\hypersetup{colorlinks=true, linkcolor=blue, citecolor=blue, linktoc=page}
\usepackage{indentfirst}
\usepackage[final]{graphicx}
\graphicspath{ {./images/} }
\usepackage{ragged2e}
\usepackage{titlesec}
\usepackage{cancel}
\usepackage{enumitem}
\usepackage{caption}
\usepackage{breqn}
\usepackage{braket}
\usepackage{multirow}
\usepackage{siunitx}
\usepackage{soul}
\usepackage[dvipsnames]{xcolor}

\makeatletter
\let\cat@comma@active\@empty
\makeatother

\usepackage[labelfont=bf, justification=Justified, format=plain]{caption}

\setcitestyle{numbers,open={[},close={]}} 

\def\cF{{\cal F}}
\def\bE{{\overline{E}}}

\newcommand{\bx}{\overline{x}}

\newcommand{\bU}{\overline{U}}
\newcommand{\bom}{\overline{\Omega}}
\newcommand{\bPhi}{\overline{\Phi}}
\newcommand{\bphi}{\overline{\phi}}
\newcommand{\bbE}{\overline{E}}
\newcommand{\cE}{{\cal E}}
\newcommand{\cV}{{\cal V}}
\newcommand{\cH}{{\cal H}}

\newcommand{\cK}{{\cal K}}
\newcommand{\cN}{{\cal N}}
\newcommand{\cO}{{\cal O}}
\newcommand{\cX}{{\cal X}}

\newcommand{\ha}{{\hat a}}

\newcommand{\hn}{{\hat n}}
\newcommand{\hH}{{\hat H}}
\newcommand{\hV}{{\hat V}}

\begin{document}
\title{Many-body tunneling in a double-well potential}

\author{M. Zendra}
\email[Corresponding author: ]{matteo.zendra@unicatt.it}
\affiliation{Dipartimento di Matematica e Fisica and Interdisciplinary Laboratories for Advanced Materials Physics, Universit{\`a} Cattolica del Sacro Cuore, via della Garzetta 48, 25133 Brescia, Italy}
\affiliation{Institute for Theoretical Physics, KU Leuven, Celestijnenlaan 200D, B-3001 Leuven, Belgium}
\affiliation{Istituto Nazionale di Fisica Nucleare, Sezione di Milano, via Celoria 16, I-20133 Milano, Italy}
\author{F. Borgonovi}
\affiliation{Dipartimento di Matematica e Fisica and Interdisciplinary Laboratories for Advanced Materials Physics, Universit{\`a} Cattolica del Sacro Cuore, via della Garzetta 48, 25133 Brescia, Italy}
\affiliation{Istituto Nazionale di Fisica Nucleare, Sezione di Milano, via Celoria 16, I-20133 Milano, Italy}
\author{G. L. Celardo}
\affiliation{Dipartimento di Fisica e Astronomia, Universit{\`a} di Firenze, via Sansone 1, 50019 Sesto Fiorentino, Firenze, Italy}
\affiliation{Istituto Nazionale di Fisica Nucleare, Sezione di Firenze, via Bruno Rossi 1, 50019 Sesto Fiorentino, Firenze, Italy} 
\author{S. Gurvitz}
\email[Corresponding author: ]{shmuel.gurvitz@weizmann.ac.il}
\affiliation{Department of Particle Physics and Astrophysics, Weizmann Institute of Science, 76100 Rehovot, Israel}
\date{\today}
\setlength{\parindent}{20pt}

\begin{abstract}
We present an approach for evaluating Wannier functions, offering an alternative perspective on their role in many-body systems. Unlike traditional methods, such as the maximally localized Wannier functions approach, which focuses on minimizing the function tails, our approach emphasizes these tails. Using perturbative analytical approximations and extensive numerical simulations on an exactly solvable model, we address nonstandard Hubbard terms and demonstrate their critical influence on many-body dynamics. Specifically, we study tunneling dynamics in arbitrary double-well potentials, moving beyond the standard Hubbard model to include nonstandard terms such as density-induced tunneling and pair tunneling. Our results reveal that these terms significantly modify the dynamics predicted by the standard Hubbard model: density-induced tunneling modifies the single-particle tunneling parameter $\Omega_0$, while pair tunneling enables coherent propagation not captured by the standard model. We show that the discrepancies between the standard and nonstandard Hubbard models grow with increasing interaction strength, potentially leading to novel transport behaviors. However, at lower interaction strengths, both models converge, as nonstandard terms become negligible. These findings have important implications for phenomena such as superconductivity in twisted bilayer graphene and metal-insulator transitions. Our model aligns well with numerical simulations of lowest-band parameters and is strongly supported by experimental observations of second-order atom tunneling in optical double-well potentials. This strong agreement with experimental data highlights the accuracy and potential of our approach in providing a more comprehensive framework for describing complex many-body systems than the standard Hubbard model.
\end{abstract}

\maketitle
\section{INTRODUCTION}
\label{sec:INTRO}
In the field of condensed-matter physics, the Hubbard model \cite{hubbard:electrons_corr, gutzwiller:ferro, essler:1dhubbard, held:electronic} serves as a foundational paradigm, crucial for understanding the behavior of various solid-state systems, especially those exhibiting strong electron correlations \cite{dutta:non_standard}. This model is characterized by a tight-binding Hamiltonian featuring single-particle tunneling energy ($\Omega_0$) allowing for hopping between neighboring sites, and an on-site two-particle interaction energy ($U$), which can be either attractive ($U<0$) or repulsive ($U>0$). Despite its apparent simplicity, the Hubbard model is widely believed to be instrumental in addressing the unresolved issues of high-$T_C$ superconductivity \cite{jiang:superconductivity, doping:lee, dagotto:highTc, bednorz:possible_high_Tc, cao:corr_insulator, anderson:resonating, brinkman:electronholecouplingcuprates, Lee:2007, Yanagisawa:2008_high_tc_superc, Haule:2007_plaquette, Gull:2012, honerkamp_dwcp, bednorz_perovskite, raghu:supuercond, licht:antiferro, sigrisr:phenom, zaanen_band_gaps, ogata:tj_model} and strongly correlated electron systems \cite{arovas:hubbard_model, andolina:nogotheorem, andolina:varyingemfield, mazza:mottgapcollapse, kunes:excitoniccondensation, miyasaka:anisotropymottgap, georges:dynamical, freericks:exact_DMFT, aoki:neqdmft, oka:photovoltaic}. However, questions have emerged regarding the model's effectiveness and empirical validation, especially in the context of interacting particles ($U \neq 0$). In fact, in the noninteracting limit, the tunneling energy $\Omega_0$ accurately describes single-particle and noninteracting many-body dynamics. However, in interacting systems, the standard Hubbard model often fails to fully capture the effects induced by interparticle interactions \cite{dutta:non_standard, jurgensen:density_induced_optical, jurgensen:observation_density_induced_tunnelling, maik:density_ind, hirsch:inapplicability}. Specifically, crucial additional terms such as \textit{density-induced tunneling} (DT) and \textit{pair tunneling} (PT) are often neglected, limiting the model's accuracy in describing real-world systems \cite{dutta:non_standard, PRB_nonstandard, buchleitner:tunn_decay, gurvitz:twoelectroncorrelated, petrosyan:liquid, terashige:doublonholonpairing, hou:manybody_tunnelling, gong:mb_destruc_tunn}.

Recent research has therefore focused on extending the Hubbard model to include these additional interaction terms, leading to what are known as nonstandard Hubbard models \cite{jurgensen:observation_density_induced_tunnelling,dutta:non_standard,photon_assisted_tunn:ma, floq_eng_corr_tunn:meinert, enh_sign_change:gorg, hirsch:electron_hole, kremer:interaction_induced_effects, zollner:tunn_dynamics, Lagoin:2022_extended_bose_hubbard}. On the one hand, when interactions between different sites of a lattice are incorporated, we refer to the \textit{extended} Hubbard model. Besides this, the DT term accounts for modifications of single-particle tunneling energy $\Omega_0$ due to the effective mean field created by other particles in the system, often referred to as \textit{bond-charge interaction} \cite{dutta:non_standard, jurgensen:density_induced_optical, jurgensen:observation_density_induced_tunnelling,supercond_state:hirsch, hubbard_nn_bc:strack, ferm_corr_hopp:karnaukhov, eta_pairing:deboer, weak_coupl_hubb:japaridze, two_point_entangl:anfossi, quantum_phase_diagr:dobry, interact_top_ins:montorsi, strack:hubbard_bond_charge, hirsch:hole_superconductivity, incomm_unconv:aligia}. On the other hand, the PT term, analogous to Cooper pair tunneling, plays a crucial role in the two-particle elastic tunneling, representing a coherent process \cite{petrosyan:liquid, kim:cooper_pair_wavefunction}.

Previous studies typically describe \textit{cotunneling} within the framework of the standard Hubbard Hamiltonian \cite{winkler:repulsively_bound_states, bloch:direct_observation, mondal:two_body_repulsive}, where it manifests as a second-order process in $\Omega_0$, generated by two virtual sequential single-particle tunnelings. Each tunneling step is characterized by a large interaction energy $|U| \gg \Omega_0$, resulting in an $\cO(\Omega_0^2/|U|)$ cotunneling frequency \cite{bloch:direct_observation}. However, the standard Hubbard model does not fully capture all relevant physical processes in interacting many-body systems. To address these limitations, different extensions to the standard Hubbard model have been investigated in Refs.~\cite{hirsch:electron_hole, amadon_hirsch:metallic, weckesser:extbose, knothe2024extended, affleck:heishubb, gilmutdinov:interplayext, chen:superconducting, adebanjo:ubiquity, kundu2023cdmfthfd, wrz:nonmonotonic, frey:hilbertspace, nicokatz:mbl, dutta:occupation, sowinski:dipolar_molecules}, incorporating DT and PT as nonstandard processes. These studies have highlighted the crucial influence of these nonstandard Hubbard terms on the behavior of strongly correlated systems. However, it is only in recent years that nonstandard Hubbard models have attracted significant attention, particularly in the context of ultracold atoms in optical lattices, where the DT phenomenon has been experimentally observed \cite{jurgensen:observation_density_induced_tunnelling, dutta:non_standard, manybody_ultracold, ultracold_mimick, hild_spin_transport, meinert_quantum, dutta:occupation, greiner:pt_mott, jaksch:cold_atom, Bohrdt:exploration_doped, trotzky:bloch_superexchange, schafer:2020, gross:quantum_sim}.

Despite these advances, fully understanding the impact of nonstandard terms on the dynamics of correlated systems remains challenging \cite{hirsch:electron_hole,luhmann:multiorb_dens_ind,dutta:non_standard}. In particular, the magnitude of these terms is intricately linked to the overlap of Wannier functions (WFs) between neighboring wells \cite{PRB_nonstandard, lihm:wannier_function, buchleitner:tunn_decay}. Current approximations often fail to account for the tails of these functions, which are crucial for determining the value and sign of nonstandard interaction matrix elements. State-of-the-art approaches to WF construction, such as the theory of maximally localized Wannier functions (MLWFs) \cite{marzari:wannier}, involve an iterative procedure to minimize the spread of WFs, making them as localized as possible in real space. While highly effective for many applications, such as the study of electronic structure in crystalline solids, transport properties, and entanglement in condensed-matter systems, MLWF's approach often neglects the long tails of the WFs, which are essential for accurately capturing the nonstandard Hubbard terms. Therefore, a consistent approximation for the WFs spanning the entire double-well region is essential for a deeper understanding of the intricate interplay between interaction and tunneling dynamics in quantum systems.

To address these limitations, we propose a novel method for evaluating WFs that spans the entire double-well region, offering a more consistent approximation. This method includes the nonstandard terms in the standard Hubbard Hamiltonian, thus reconsidering the two-particle tunneling process. Our analysis suggests that the coherent PT amplitude remains significant even when $U$ is large, potentially dominating the cotunneling process under specific system's parameters and interaction strengths \cite{gurvitz:twoelectroncorrelated}. Therefore, in addition to the standard Hubbard model, we introduce a single-band (SB) nonstandard model that exactly accounts for all interaction effects, still neglecting the effects of higher energy bands. Furthermore, we evaluate the system's dynamics using a multiband (MB) nonstandard model, which includes both the nonstandard Hubbard terms and the effects of higher energy bands in the Hubbard Hamiltonian.

After proposing our method for the evaluation of the WFs, it is essential to discuss the specific systems and models where this approach is applied. Our analysis mainly focuses on two different potentials: a square double-well potential and a sinusoidal potential. These models provide valuable insights into the dynamics of strongly correlated systems, with each model serving a different role in theoretical and experimental analysis.

Specifically, the square double-well potential is a toy-model, providing a simplified yet effective framework for understanding the main aspects of tunneling interacting dynamics, see Ref.~\cite{PRB_nonstandard}. This model allows us also to explore the basic principles without the complexities introduced by more realistic potentials. Our analytical approximated results for this model exhibit excellent agreement with our numerical simulations, demonstrating the consistency, validity and reliability of our approach.

On the other hand, the sinusoidal potential represents a more realistic model that closely mimics experimental conditions, particularly in systems such as optical lattices. Our model's predictions for the sinusoidal potential show excellent agreement with experimental data in Ref.~\cite{bloch:direct_observation}. Specifically, compared with the standard Hubbard model and the SB nonstandard model, the MB nonstandard model, which incorporates the nonstandard Hubbard terms and the effects of higher-energy bands in the Hubbard Hamiltonian, provides a more accurate alignment with experimental results. The MB nonstandard model effectively describes the two-particle tunneling process, both for weakly and strongly interacting regimes, emerging as the most accurate representation of experimental observations.

In general, the nonstandard Hubbard model diverges significantly from the standard Hubbard model as the interaction strength increases, leading to different transport regimes. In the nonstandard model, strong interactions modify single-particle tunneling while enhancing PT, creating an interplay between these effects and potentially leading to new transport phenomena beyond the metal-insulator transitions \cite{anderson:valence, imada:review, Mott:basis}. However, at lower interaction strengths, both models yield similar results, particularly when DT and PT terms are negligible compared with $\Omega_0$.

Finally, our approach allows for further modifications to the potential shape, enhancing its flexibility and applicability. Specifically, we adapt our method to obtain analytical approximated results for a squared cosine potential, that exhibit excellent agreement with established theoretical results available in the literature, see Ref.~\cite{luhmann:multiorb_dens_ind}.

In summary, our model's ability to consistently approximate WFs across the entire double-well region and its successful application to both toy and realistic potentials underscore its robustness and versatility. The strong correlation between our analytical approximated, numerical, and experimental results confirms that our method provides a comprehensive and accurate description of tunneling dynamics and interactions in quantum systems.

After recalling the two-potential approach (TPA) as discussed in Ref.~\cite{gurvitz:twopotapproach}, we accurately define the WFs of a double-well potential in Sec.~\ref{sec:TPA}. In Sec.~\ref{sec:NSHM}, we analyze the nonstandard Hubbard model, evaluating the corresponding nonstandard Hubbard terms through the TPA for a contact interaction in Sec.~\ref{sec:CONTACT_INT}. Note that Sec.~\ref{sec:TPA} and \ref{sec:NSHM} contain material from our previous work, see Ref.~\cite{PRB_nonstandard}. We compare the results with existing theoretical literature in Sec.~\ref{sec:COMP_THEOR}. Finally, in Sec.~\ref{sec:DYNAMICS}, we examine the effects of DT and PT terms on the dynamics of two distinguishable particles in a square double-well potential. Our findings are compared with experimental observations in Sec.~\ref{sec:COMP_EXP}, highlighting the regimes where nonstandard DT and PT terms are particularly relevant.

\section{TWO-POTENTIAL APPROACH AND WANNIER FUNCTIONS FOR A DOUBLE-WELL POTENTIAL}
\label{sec:TPA}
In this section, we employ an analytical approach, see Ref.~\cite{PRB_nonstandard}, for the evaluation of the WFs within the framework of a symmetric square double-well potential. It is based on the modification of the orbital wave functions (hereinafter orbitals), by incorporating their extensions into neighboring sites. It derives from the two-potential approach (TPA) to tunneling problems, initially developed for tunneling to the continuum \cite{gurvitz:decay_width, gurvitz:novel_approach, gurvitz:modified_pot_approach}, and it has been adapted to compute the discrete eigenspectrum of a multiwell system. The corresponding set of eigenstates, obtained from the modified orbitals, is used to determine the WFs through a proper unitary transformation. Consequently, this approach avoids the ambiguities associated with approaches based on the continuous Bloch-function spectrum \cite{marzari:wannier}.

For simplicity, we start presenting our method for a generic symmetric double-well potential, although the method can straightforwardly be extended to multiwell systems, see Supplemental Material of Ref.~\cite{PRB_nonstandard}, to validate its accuracy. We then compare our results for the WFs with the orbitals and the exact solutions of the Schr\"odinger equation for the case of two distinguishable particles. Finally, the resulting WFs can be exploited to correctly evaluate the DT and PT terms of the nonstandard Hubbard Hamiltonian. 

Initially, consider a particle placed in a generic one-dimensional (1D) double-well potential formed by two single-well potentials $\cV_1(x)$ and $\cV_2(x)$, separated by a potential barrier. To simplify the analytical treatment, we assume a symmetric double-well potential, even if the procedure can be easily extended to generic asymmetric potentials. The single-well potentials are defined such that
$$\cV_1(x)=\cV_2(-x)$$
for $x>x_0\,,$ and
$$\cV_{1,2}(x_0)=const\,.$$
It is important to note that this procedure differs significantly from the widely used Wigner $R$-matrix theory (see Ref.~\cite{lane:rmatrix}), in which the single-well potential is truncated, and an infinite potential barrier is placed beyond the separation point.

\begin{figure*}[t]
\centering
{\includegraphics[width=17.2cm]{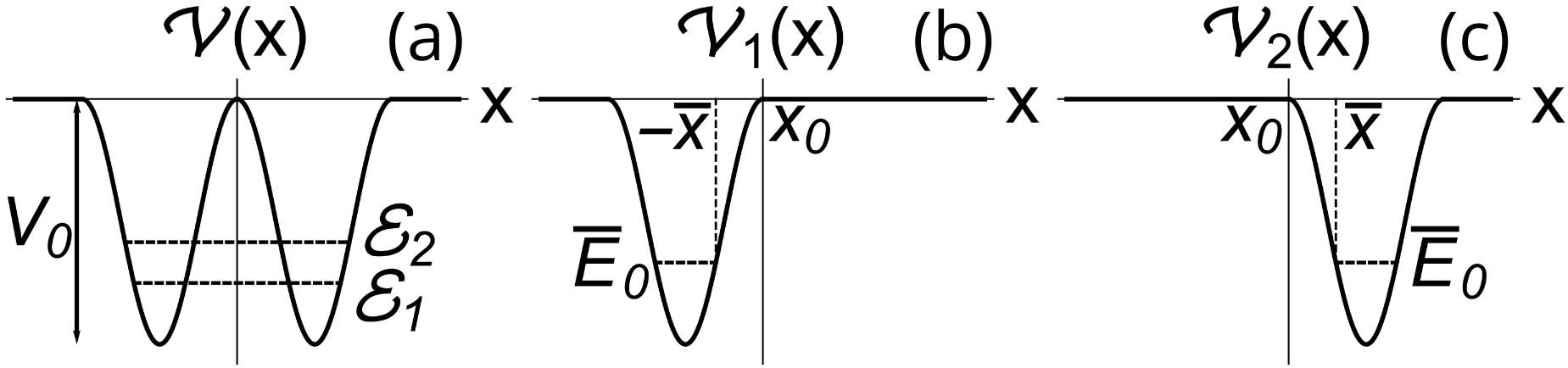}}
\caption{(a) Symmetric double-well potential $\cV(x)$, given by the sum of two single-well potentials, with lattice depth $V_0$. Dashed lines represent the two lowest-band eigenstates of the system, $\cE_1$ and $\cE_2$. (b), (c) Left and right single-well potentials $\cV_1(x)$ and $\cV_2(x)$, where $\mp \,\bx$ are the classical turning points, so that $\cV_1\left(\mp \,\bx \right)=\bE_0$ and $\cV_2\left(\mp \,\bx \right)=\bE_0$, and $x_0$ is the separation point. Each of the single-well potentials vanishes beyond the separation point ($x_0=0$), defined so that $\cV_1(x_0)=\cV_2(x_0)=0$. Each of the single-well potentials contains one bound state with energy $\bE_{0}<0$.}
\label{fig1_dw}
\end{figure*}

By choosing the energy scale such that $\cV(x_0)=0$, we can express the potential as
\begin{equation}
\cV(x)=\cV_1(x)+\cV_2(x)\,,
\label{2_generic_well}
\end{equation}
as shown in Fig.~\ref{fig1_dw} for $x_0=0$. If $\cV(x_0)\not =0$, then $\cV(x_0)$ must be subtracted in Eq.~\eqref{2_generic_well} (see Ref.~\cite{gurvitz:novel_approach}). Importantly, the final result is independent of the chosen energy scale. A similar construction for single-well potentials can be applied to the extension of the TPA to the multidimensional case, see Ref.~\cite{gurvitz:twopotapproach}, where the separation point is replaced by a separation surface, beyond which the corresponding single-well potential becomes constant.

To simplify our analysis and in close analogy with the Hubbard model, we employ the SB approximation, by focusing on the lowest energy band and neglecting all the others, as well as the continuous spectrum ($E\ge 0$). We then assume that the two lowest bound states of the system (with energy $\cE_1<0$ and $\cE_2<0$) create a band, well separated from the others. The eigenstates belonging to this band, namely,
$$\psi_{1,2}(x)= \braket{x|\psi_{1,2}}\,,$$ are obtained from the exact solution of the Schr\"odinger equation
\begin{equation}
\cH\ket{\psi_{1,2}}\equiv \left(\cK+\cV(x)\right)\ket{\psi_{1,2}}=\cE_{1,2}\ket{\psi_{1,2}}\,,
\label{Schr_eq_k}
\end{equation}
where
$$\cK=-\nabla_x^2$$
is the kinetic term, and $\cV(x)$ the double-well potential in Eq.~\eqref{2_generic_well}
(to simplify the notation, we use dimensionless units, i.e., $\hbar=2m=1$, unless otherwise specified).

The tight-binding tunneling Hamiltonian $H$ describing this lowest band is given by
\begin{equation}
H=\bbE_0\sum_{j=1}^{2}\ket{\Psi_j}\bra{\Psi_j}+\bom_0\left(\ket{\Psi_1}\bra{\Psi_2}+H.c.\right)\,,
\label{tunn_ham}
\end{equation}
where $\bE_0$ is the single-well energy, see Fig.~\ref{fig1_dw}, $\bom_0$ is the nearest-neighbor tunneling coupling, while $$\Psi_j(x)=\braket{x|\Psi_j}$$ is the $j$-th well WF. The site energies $\bE_0$ and the WFs $\Psi_j(x)$ can be obtained by diagonalizing the tunneling Hamiltonian in Eq.~\eqref{tunn_ham}. By identifying the eigenvalues and the eigenstates of the tunneling Hamiltonian in Eq.~\eqref{tunn_ham} respectively with the exact energies $\cE_{1,2}$ and wave functions $\ket{\psi_{1,2}}$ of the double-well system obtained from Eqs.~\eqref{Schr_eq_k}, we can write
\begin{subequations}
\begin{align}
\bbE_0&=\left(\cE_1+\cE_2\right)/2\,,
\label{Wannier_functions_A}\\
\bom_0&=(\cE_1-\cE_2)/2\,,
\label{Wannier_functions_B}\\
\Psi_{L,R}(x)&=\frac{1}{\sqrt{2}}\left[\psi_1(x)\pm\psi_2(x)\right]\,,
\label{Wannier_functions_C}
\end{align}
\end{subequations}
where $\bbE_0$ and $\bom_0$ are the tunneling Hamiltonian parameters, while $$\Psi_{L,R}(x)=\braket{x|\ha^\dagger_{L,R}|0}$$ are the WFs, where $\ha^\dagger_{L,R}$ are the creation operators for the left and right well, respectively.

In the multiwell limit, the previous procedure resembles the standard one for constructing WFs from the continuous spectrum of Bloch functions under periodic boundary conditions. In that case, due to the phase indeterminacy \cite{marzari:wannier}, the resulting WFs are not uniquely defined. In contrast, we impose the boundary conditions
$$\psi_{1,2}(x)\sim e^{-\sqrt{-\cE_{1,2}}|x|}$$
for $x\to\pm\infty$ on the eigenstates $\psi_{1,2}(x)$ belonging to the first band. This ensures that the solution to the Schr\"odinger equation \eqref{Schr_eq_k} is uniquely defined for both the energies $\cE_{1,2}$.

Typically, determining the spectrum of an arbitrary double-well potential system involves exact numerical diagonalization. Consequently, many studies use an approximate form for the WFs $\Psi_{L,R}(x)$. In contrast, we evaluate the WFs by modifying in a perturbative way the single-well orbitals. To do so, we define the left- and right-well orbitals as $$\Phi_0^{(1,2)}(x)\equiv\Phi_0^{(1,2)}(E_0,x)\,,$$ where $E_0$ is the ground-state energy of the single-well potential. Therefore, the orbitals $\Phi_0^{(1,2)}(x)$ are respectively the ground states of the left- and right-well Hamiltonian
\begin{equation}
H_{1,2}\Phi_0^{(1,2)}(x)=E_0\Phi_0^{(1,2)}(x)\,,
\label{orbitals_N2}
\end{equation}
where $$H_{1,2}=\cK+\cV_{1,2}(x)$$ are the Hamiltonians of the left and right well, respectively. Note that $\Phi_0^{(2)}(x)=\Phi_0^{(1)}(-x)$ due to the symmetry of the double-well system. Since each potential $\cV_{1,2}(x)$ exactly vanishes beyond the separation point $x_0=0$ (see Fig.~\ref{fig1_dw}), the orbitals can be written as
\begin{subequations}
\label{orbitals_0}
\begin{align}
&\Phi_0^{(1)}(x)=\Phi_0^{(1)}(0)e^{-q_0x}\quad&\text{for $x\ge 0$}\,,\\
&\Phi_0^{(2)}(x)=\Phi_0^{(2)}(0) e^{q_0x}\quad&\text{for $x\le 0$}\,,
\end{align}
\end{subequations}
where $q_0=\sqrt{-E_0}$. Since $\cV_{1,2}(x)\to 0$ in the limit $x\to\pm\infty$, the solution of the Schr\"odinger equation within the potential barrier ($E<0$) is a combination of two functions $\approx e^{\pm q_0x}$. Disregarding the growing solution as unphysical, the decreasing one uniquely establishes the allowed discrete energy levels. Conversely, beyond the potential barrier ($E\ge 0$), the exponential factor $q_0$ becomes imaginary, indicating the presence of both the solutions and a resulting continuous spectrum.

As said, our approach consists in using the orbitals $\Phi_0^{(1,2)}(x)$ as a basis for constructing the eigenstates and the corresponding WFs in a perturbative way. A possible approach would be to consider the left-well orbital $\Phi_0^{(1)}(x)$ as an unperturbed eigenstate of the Hamiltonian $H_1$, and the right-well potential $\cV_2(x)$ as the perturbation (or vice versa). However, this approach would lack a perturbative expansion parameter. In contrast, the problem can be solved using the TPA method, by employing the overlap $$\beta \equiv \braket{\Phi_0^{(1)} |\Phi_0^{(2)}}$$ as a perturbative expansion parameter. We note that $\beta$ can be considered as a small parameter, since it is of the order of the barrier penetration coefficient $$T_0=\exp \left(-\int_{-\bx}^{\bx}|q(x')|\,dx'\right)\ll 1,$$ where $q(x)$ is the (imaginary) momentum under the barrier and $\pm \,\bx$ are the classical turning points, indicated in Fig.~\ref{fig1_dw}. Using this approach, we obtain for the tunneling Hamiltonian parameters in Eqs.~\eqref{Wannier_functions_A} and \eqref{Wannier_functions_B}:
\begin{subequations}
\label{params_Ham}
\begin{align}
&\bE_0=E_0+\cO(\beta^2)\,,\\
&\bom_0=\Omega_0+\cO(\beta^2)\,,
\end{align}
\end{subequations}
where the site energy $E_0$ is given by Eqs.~\eqref{orbitals_N2}. Similarly,
$$\cE_{1,2}=E_\pm +\cO(\beta^2)\,,$$
where $$E_\pm=E_0\pm\Omega_0\,.$$ Thus, the parameters of the tunneling Hamiltonian in Eq.~\eqref{tunn_ham} are completely determined by the single-well orbitals in Eqs.~\eqref{orbitals_N2}. Specifically, the tunneling coupling $\Omega_{0}$ is defined as
\begin{equation}
\Omega_{0}=\braket{\Phi_0^{(1)}|\cV_2|\Phi_0^{(2)}}=\braket{\Phi_0^{(1)}|\cV_1|\Phi_0^{(2)}}\,.
\label{tcoup}
\end{equation}
Note that $\Omega_0$ is proportional to the barrier tunneling penetration coefficient $T_0$, making it a small parameter in the corresponding perturbative expansion of the eigenenergies. From Eq.~\eqref{tcoup}, it follows that $\Omega_0$ is always negative, since the orbitals $\Phi_0^{(1,2)}(x)$ correspond to the ground states of the respective wells (and their product is positive), while the potentials $\cV_{1,2}(x)<0$, as shown in Fig.~\ref{fig1_dw}. From the Schr\"odinger equation, Eq.~\eqref{tcoup} can be rewritten as
\begin{equation}
\label{bardeen}
\Omega_{0}=\Phi_0^{(1)\prime}(0)\Phi_0^{(2)}(0)-\Phi_0^{(1)}(0)\Phi_0^{(2)\prime}(0)\,.
\end{equation}
From Eqs.~\eqref{orbitals_0}, we can evaluate the derivative of the orbitals at the separation point $x_0=0$:
\begin{subequations}
\label{der_orbitals}
\begin{align}
&\Phi_0^{(1)\prime}(0)=-\sqrt{|E_0|}\Phi_0^{(1)}(0)\,,\\
&\Phi_0^{(2)\prime}(0)=\sqrt{|E_0|}\Phi_0^{(2)}(0)\,.
\end{align}
\end{subequations}
By substituting Eqs.~\eqref{der_orbitals} into Eq.~\eqref{bardeen}, we derive the tunneling energy $\Omega_0$, given by
\begin{equation}
\Omega_0=-2\sqrt{|E_0|}\Phi_0^{(1)}(0)\Phi_0^{(2)}(0)\,,
\label{omega_0_bardeen}
\end{equation}
which represents a product of neighboring orbitals evaluated at the separation point. Examining Eq.~\eqref{omega_0_bardeen}, we observe that $\Omega_0\propto T_0$ serves as a simplified one-dimensional (1D) version of the well-known Bardeen formula \cite{bardeen:tunnelling_many_body, gawelczyk:bardeen}.

As discussed in Ref.~\cite{PRB_nonstandard}, the eigenstates $$\psi_\pm(x)\equiv\psi_{1,2}(E_\pm,x)$$ cannot be obtained from Eq.~\eqref{Wannier_functions_C} by replacing the WFs $\Psi_{L,R}(x)$ with the corresponding orbitals $\Phi_0^{(1,2)}(x)$ in Eqs.~\eqref{orbitals_N2}. In fact, by doing that, an inconsistency between the energy arguments of the double-well wave functions $\psi_\pm(x)$ and the single-well orbitals $\Phi_0^{(1,2)}(x)$ arises. To overcome this problem, we introduce an energy shift ($E_\pm-E_0$) in the orbitals by replacing the ground-state energy $E_0$ with a free parameter $E<0$. The resulting modified orbitals $$\bPhi^{(1,2)}(E,x)\equiv \Phi_0^{(1,2)}(E_0\to E,x)$$ (normalized to unity) are obtained from Eqs.~\eqref{orbitals_N2} by imposing the boundary conditions
$$\bPhi^{(1,2)}(E,x\to \mp\infty)\propto e^{\pm q x}\,,$$
where $q=\sqrt{-E}$ (compare with Eqs.~\eqref{orbitals_0}). These boundary conditions uniquely define the modified orbitals for any energy value $E$. Unlike $\Phi_0^{(1,2)}(x)$, the modified orbitals $\bPhi^{(1,2)}(E,x)$ asymptotically diverge for any $E\not =E_0$. Thus, we define them on two different segments, respectively $\cX_1=(-\infty, 0)$ and $\cX_2=(0,\infty)$, vanishing elsewhere. As a result, they are {\em nonoverlapping}, and therefore {\em orthogonal}. The eigenenergies $E=E_{\pm}$ are determined by enforcing continuity conditions at the separation point $(x_0=0)$. The corresponding eigenstates, $\psi_\pm(x)$, are expressed in terms of the modified orbitals $\bPhi^{(1,2)}(E,x)$, which are defined within the segment of the $j$-th well $(j=1,2)$, as follows:
\begin{equation}
\psi_\pm(x)=\frac{1}{\sqrt{2}}\left[\bPhi^{(1)}(E_\pm,x)\pm\bPhi^{(2)}(E_\pm,x)\right]\,.
\label{eigen_site}
\end{equation}
This approach constructs the exact eigenspectrum, $\cE_{1,2}\equiv E_\pm$ and $\psi_{1,2}(x)\equiv\psi_\pm(E_\pm,x)$, by using the modified orbitals for different segments and imposing continuity at the separation point $(x_0=0)$. This method represents the standard procedure for solving the Schr\"odinger equation \eqref{Schr_eq_k}. The resulting WFs, given by Eqs.~\eqref{Wannier_functions_C}, are expressed as
\begin{equation}
\begin{aligned}
&\Psi_{L,R}(x)=\frac{1}{\sqrt{2}}\left[\psi_+(E_+,x)\pm\psi_-(E_-,x)\right]=\\
&=\frac{1}{2}\left[\bPhi^{(1)}(E_+,x)+\bPhi^{(2)}(E_+,x)\right]+\\
&\pm\frac{1}{2}\left[\bPhi^{(1)}(E_-,x)-\bPhi^{(2)}(E_-,x)\right]\,.
\end{aligned}
\label{psi_lr}
\end{equation}
These WFs are uniquely defined by the modified orbitals $\bPhi^{(1,2)}(E,x)$, thereby avoiding the phase indeterminacy present in standard approaches (see Ref.~\cite{marzari:wannier}). Furthermore, the electron dynamics described by the tunneling Hamiltonian $H$ in Eq.~\eqref{tunn_ham} using these WFs is equivalent to that derived from the original Schr\"odinger Hamiltonian \eqref{Schr_eq_k} in the single-band approximation. This equivalence represents a crucial criterion for defining WFs.

It is noteworthy that $\Psi_{L,R}(x)$ in Eq.~\eqref{psi_lr} and their derivatives are continuous over the entire interval $-\infty<x<\infty$ due to continuity of the eigenfunctions $\psi_\pm(E_\pm,x)$ and their derivatives. Consequently, the representation of the WFs in terms of the modified orbitals $\bPhi^{(1,2)}$ must also be continuous, even though each orbital may exhibit discontinuities at the separation point $x=0$.

Equations.~\eqref{psi_lr} can be significantly simplified by employing a perturbative expansion within the TPA framework to approximate the energy spectrum in terms of the tunneling coupling $\Omega_0$, since $\Omega_0 \propto \beta \propto T_0$. By neglecting terms involving $\Omega_0^2$, the eigenenergies can be expressed using single-well orbitals, exploiting
$$\cE_{1,2}\simeq E_0\pm\Omega_0\,,$$
where $E_0$ is derived from Eqs.~\eqref{orbitals_N2} and the energy shift $\Omega_0$ is determined by the Bardeen formula in Eq.~\eqref{omega_0_bardeen}. Expanding the modified orbitals $\bPhi^{(1,2)}(E,x)$ in Eqs.~\eqref{psi_lr} in powers of $E_\pm-E_0=\pm\Omega_0$, and neglecting $\cO(\Omega_0^2$) terms, the WFs can be approximated as:
\begin{subequations}
\label{wan_approx}
\begin{align}
\Psi_L(x)=\bPhi_0^{(1)}(x) 
+\Omega_0\partial_E\bPhi_0^{(2)}(x)\,,
\label{wan_approx_A}\\
\Psi_R(x)=\bPhi_0^{(2)}(x)
+\Omega_0\partial_E\bPhi_0^{(1)}(x)\,.
\label{wan_approx_B}
\end{align}
\end{subequations}
This result, first derived in Ref.~\cite{PRB_nonstandard} for a triple-well potential, can be further extended to multi-well systems. Here, $$\bPhi_0^{(1)}(x)\equiv \Phi_0^{(1)}(E_0,x)\,, \quad \bPhi_0^{(2)}(x)\equiv \Phi_0^{(2)}(E_0,x)$$ for $x\in (\cX_1,\cX_2)$ (up to negligible corrections to the normalization) and vanish elsewhere, while $$\partial_E\bPhi_0^{(1,2)}(x) \equiv [\partial\bPhi^{(1,2)}(E,x)/\partial E]_{E\to E_0}\,.$$ Note that due to symmetry, $\bPhi_0^{(2)}(x)=\bPhi_0^{(1)}(-x)$. The resulting WFs $\Psi_{L,R}(x)$ in Eqs.~\eqref{wan_approx} are already orthogonal, so that no more orthogonalization of the orbitals is needed. Equations.~\eqref{wan_approx} are composed of two {\em nonoverlapping} terms, that describe the WFs within their respective wells and their tails extending into adjacent wells. The second term, which is the contribution related to the WFs's tails, is proportional to $\Omega_0$, and it is significantly smaller than the first one. This distinction highlights the close relation between the WFs' tails and the tunneling to neighboring sites, making them clearly different from corresponding orbitals. We emphasize that the WFs $\Psi_{L,R}(x)$ and their derivatives remain continuous at $x=0$, up to $\cO(\Omega_0^2$) terms, as implied by Eqs.~\eqref{psi_lr}. In fact, unlike the nodeless orbitals, the WFs' tails change sign, aligning with the orthogonality of $\Psi_{L,R}(x)$.
\begin{figure}[t]
\centering
\includegraphics[width=8.6cm]{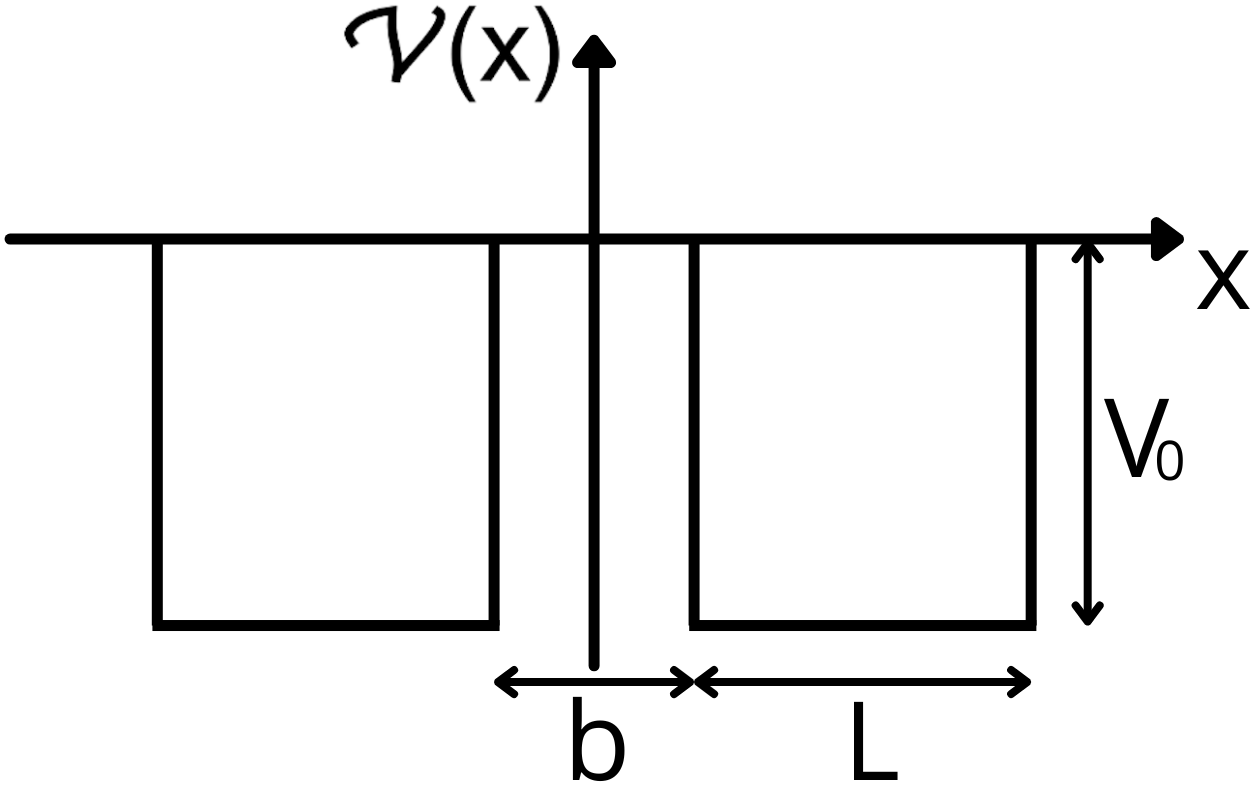}
\caption{Symmetric square double-well potential, consisting of two rectangular wells of width $L$ separated by a barrier of width $b$, with lattice depth $V_0>0$. Unless otherwise specified, we consider only one energy level per each well (SB approximation), discarding all the other bound states, as well as the continuous spectrum.}
\label{fig3_sdw}
\end{figure}
We exactly evaluate the WFs $\Psi_j(x)$ of the square double-well potential, shown in Fig.~\ref{fig3_sdw}, comparing the results with those obtained with the TPA, and with the corresponding orbitals. The potential $\cV(x)$ consists of two coupled rectangular potential wells, labeled by $j=1,2\equiv L,R$, respectively denoting left and right well, given by
\begin{equation}
\cV(x)=-V_0 \quad {\rm for}\quad \frac{b}{2}<|x|<L+\frac{b}{2}\,.
\label{sqdouble}
\end{equation}
From Eqs.~\eqref{eigen_site}, we obtain the exact eigenstates in terms of the modified orbitals $\bPhi^{(1,2)}(E_\pm,x)$, where $E_\pm$ correspond to the ground state and excited state, respectively. Since the modified orbitals $\bPhi^{(1,2)}(E,x)$ are normalized to unity for any energy $E$, we can demonstrate the orthogonality of the WFs in Eqs.~\eqref{wan_approx} explicitly. Since the normalization factor of the modified orbitals $\bPhi^{(1,2)}(E,x)$ is energy dependent, it is useful to highlight it by introducing the following reduced orbitals
\begin{equation}
\bPhi^{(1,2)}(E_\pm,x)\equiv \cN_\pm\bphi^{(1,2)}(E_\pm,x)\,,
\label{modif_orbitals}
\end{equation}
where $\cN_\pm$ is the normalization factor, defined as
\begin{equation}
\cN_{\pm}\equiv \cN(E_{\pm})=\left(\int\limits_{-\infty}^0\left[\bphi^{(1)}(E_{\pm},x)\right]^2 \,dx\right)^{-1/2}\,,
\label{norm}
\end{equation}
while
\begin{widetext}
\begin{subequations}
\label{phi_site}
\begin{align}
&\bphi^{(1)}(E,x)=
\begin{cases}
\frac{p}{\sqrt{V_0}}e^{q \left(x+L+\frac{b}{2}\right)}\,&{\rm for}\,-\infty <x<-L-\frac{b}{2}\\
\cos \left[p\left(x+\frac{L+b}{2}\right)+\varphi\right]\,&{\rm for}\,-L-\frac{b}{2}<x<-\frac{b}{2}\\
\cF_1(E)e^{-q\left(x+\frac{b}{2}\right)}+\cF_2(E)e^{q\left(x+\frac{b}{2}\right)}\,&{\rm for}\,\quad -\frac{b}{2}<x<\infty
\end{cases}\,,\\
&\bphi^{(2)}(E,x)=\bphi^{(1)}(E,-x)\,,
\end{align}
\end{subequations}
\end{widetext}
where $p\equiv p(E)=\sqrt{V_0+E}$ and $q\equiv q(E)=\sqrt{-E}$.
Here,
\begin{subequations}
\label{F1_F2_phase_segments}
\begin{align}
&\cF_1(E)=\frac{p}{\sqrt{V_0}}\cos (2\varphi)+\frac{p^2-q^2}{2q\sqrt{V_0}}\sin (2\varphi)\,,\label{F1_segments}\\
&\cF_2(E)=-\frac{\sqrt{V_0}}{2q}\sin (2\varphi)\,,\label{F2_segments}\\
&\varphi\equiv\varphi(E)=\frac{pL}{2}-\arccos \frac{p}{\sqrt{V_0}}\,,\label{phi_phase_segments}
\end{align}
\end{subequations}
with the phase $\varphi$ obtained from the matching conditions at $x=-L-b/2$. The eigenenergies are obtained by matching the logarithmic derivatives of $\bphi^{(1)}(E,x)$ and $\bphi^{(2)}(E,x)$ at $x=0$. From Eqs.~\eqref{phi_site}, we derive
\begin{equation}
\cF_1(E)=\pm\cF_2(E)e^{q b}\,.
\label{F12_segments}
\end{equation}
Solving Eq.~\eqref{F12_segments} for $E$, we find the first two eigenenergies $E_\pm$ of the double-well system. Then, substituting $$\bPhi^{(1,2)}(E_\pm,x)\equiv \bPhi_\pm^{(1,2)}(x)$$ into Eqs.~\eqref{eigen_site}, we find the exact eigenstates for the double-well potential.
\begin{figure*}[t]
\centering
{\includegraphics[width=17.2cm]{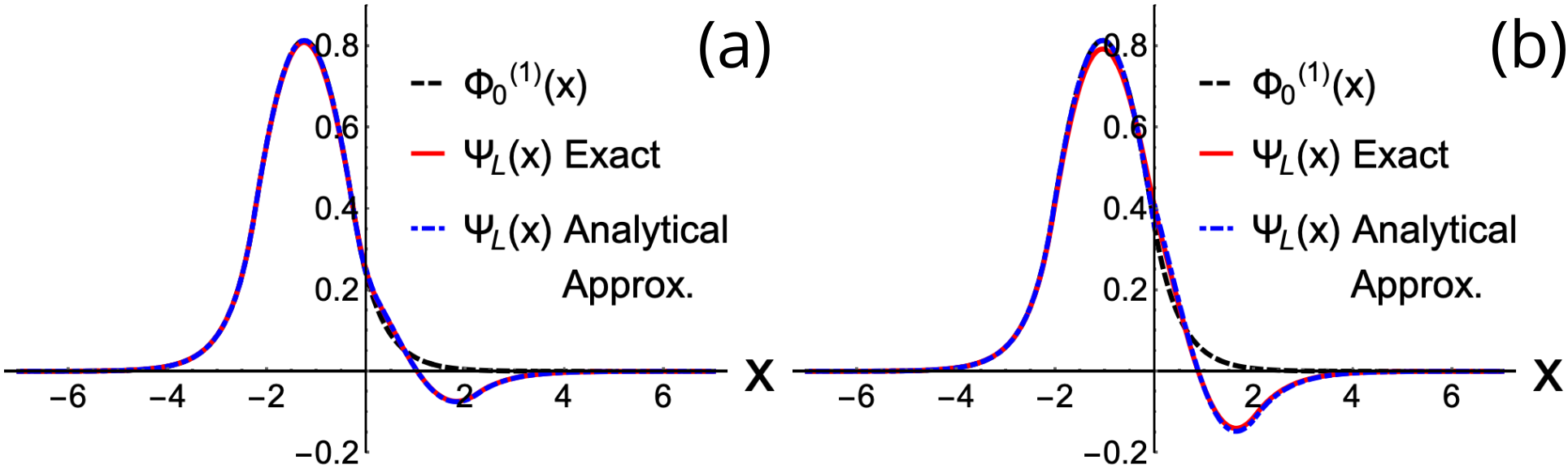}}
\caption{Comparison between the left-well orbital $\Phi_0^{(1)}(x)$ (black dashed curve), the exact (red solid curve) and our analytical approximated (blue dot-dashed curve) left-well WF $\Psi_L(x)$. The results are obtained considering the double-well potential in Fig.~\ref{fig3_sdw} for $V_0=5$, $L=2$, and two different barrier widths: $b=0.5$ in panel (a) and $b=0.1$ in panel (b). The exact ground-state energy $E_0$ of the single-well and the tunneling coupling $\Omega_0$ are $\bE_0=\frac{E_++E_-}{2} \simeq -3.866, -3.940$ and $\bom_0=\frac{E_+-E_-}{2} \simeq -0.223, -0.489$, respectively. The corresponding parameters given by the TPA (single-well) are $E_0 \simeq -3.867, -3.932$ and $\Omega_0 \simeq -0.224, -0.490$, respectively.}
\label{fig4_tails}
\end{figure*}
From Eqs.~\eqref{modif_orbitals}, it follows $$\bPhi_0^{(1,2)}(x)={\cN}_0\bphi_0^{(1,2)}(x)\,,$$ where $\bphi_0^{(1,2)}(x)\equiv \bphi_0^{(1,2)}(E_0,x)$ and $\cN_0 \equiv \cN(E_0)$ is the normalization factor, defined as:
\begin{equation}
\cN_0=\left(\frac{L}{2}+\frac{1}{q_0}\right)^{-1/2}\,.
\label{norm_factor}
\end{equation}
Expanding the term $\cN_\pm$ in Eq.~\eqref{norm} up to $\cO(\Omega_0)$ terms, we obtain
$$\cN_\pm \simeq \cN_0\pm \Omega_0(\partial_E\cN_0)\,,$$
where
\begin{equation}
\begin{aligned}
\label{der_norm}
&\partial_E \cN_0 = \left. \frac{\partial \cN (E)}{\partial E} \right |_{E\to E_0}=\\
&=-\cN_0^3 \int\limits_{-\infty}^0 \bphi^{(1)}(E,x)\left.\frac{\partial}{\partial E}\bphi^{(1)}(E,x)\right |_{E\to E_0}\,dx\,.
\end{aligned}
\end{equation}
Substituting $\bPhi_0^{(1,2)}(x)$ into Eqs.~\eqref{wan_approx}, we obtain (up to ${\cal O}(\Omega_0^2$) terms)
\begin{equation}
\begin{aligned}
\label{wan_approx_norm}
\Psi_{L,R}(x)&=\cN_0 \bphi_0^{(1,2)}(x)+\\
&+\Omega_0\left[\cN_0\partial_E\bphi_0^{(2,1)}(x)+\bphi_0^{(2,1)}(x)\left(\partial_E\cN_0\right)\right]\,,
\end{aligned}
\end{equation}
where $$\partial_E\bphi_0(x)\equiv[\partial\bphi(E,x)/\partial E]_{E\to E_0}\,.$$ From Eqs.~\eqref{norm} and Eq.~\eqref{der_norm}, we can verify that the WFs $\Psi_{L,R}(x)$ in Eqs.~\eqref{wan_approx_norm} are orthogonal (up to $\cO(\Omega_0^2)$ terms):
\begin{equation}
\begin{aligned}
&\int\limits_{-\infty}^0\Psi_L(x)\Psi_R(x)\,dx=\\
&=\Omega_0\cN_0^2\left(1-\cN_0^2
\int\limits_{-\infty}^0\left[\bphi_0^{(1)}(x)\right]^2\,dx\right)=\\
&=\int\limits_{-\infty}^0\bphi_0^{(1)}(x)\partial_E\bphi_0^{(1)}(x)\,dx=\cO(\Omega_0^2)\,.
\end{aligned}
\label{orth}
\end{equation}

Equations.~\eqref{wan_approx}, or equivalently Eqs.~\eqref{wan_approx_norm}, represent our main result regarding the WFs, that can be relevant for a broad range of multiwell systems (for further details about the application of our method to different types of multiwell systems, see Appendix \ref{appendix_A} and Appendix \ref{appendix_B}). In the following, we assess these WFs predictions by comparing them with the orbitals and the results obtained from the exact solutions of the Schr\"odinger equation, considering the symmetric double-well potential in Eq.~\eqref{sqdouble}. For the TPA results, the eigenenergies are evaluated via single-well orbitals, using Eqs.~\eqref{orbitals_N2}. To make this comparison, we consider $E=E_0\pm\Omega_0$ in Eqs.~\eqref{F1_F2_phase_segments} and \eqref{F12_segments}, together with
\begin{equation}
\begin{split}
& E_0=(E_++E_-)/2\,,\nonumber\\
& \Omega_0=(E_+-E_-)/2\,.\nonumber\\
\end{split}
\end{equation}
We solve Eq.~\eqref{F12_segments} by taking into account that the barrier penetration coefficient $e^{-qb}\propto\Omega_0$, and by using
\begin{equation}
\varphi(E_0\pm\Omega_0)=\pm\left(1+q_0\frac{L}{2}\right)\frac{\Omega_0}{2p_0q_0} \equiv \pm\frac{\Omega_0}{2p_0\cN_0^2}\,.
\label{phase}
\end{equation}
By expanding Eq.~\eqref{F1_segments} and \eqref{F2_segments} in powers of $\Omega_0$, we obtain
\begin{subequations}
\label{F12_expans_om}
\begin{align}
&\cF_1(E)\approx \frac{p_0}{\sqrt{V_0}}+\cO(\Omega_0^2)\,,\\
&\cF_2(E)\approx -\frac{\sqrt{V_0}}{q} \varphi+\cO(\Omega_0^2)\,.
\end{align}
\end{subequations}
Substituting Eqs.~\eqref{F12_expans_om} in Eq.~\eqref{F12_segments}, we obtain
\begin{equation}
\Omega_0=-2\cN_0^2\frac{q_0p_0^2}{V_0}e^{-q_0 b}\,,
\label{omega_0_tunn_an}
\end{equation}
that corresponds to the Bardeen formula in Eq.~\eqref{omega_0_bardeen} obtained with the TPA.

The exact WFs $\Psi_{L,R}(x)$ are obtained expanding the modified orbitals in Eqs.~\eqref{wan_approx_norm} up to $\cO(\Omega_0)$ terms, where
\begin{widetext}
\begin{subequations}
\label{phi_der}
\begin{align}
&\partial_E \bphi^{(1)}_0(x)=
\begin{cases}
\frac{1} {2\sqrt{V_0}}\left[\frac{1}{p_0}-\frac{p_0}{q_0}\left(x+L+\frac{b}{2}\right)\right]e^{q_0 \left(x+L+\frac{b}{2}\right)}\,&{\rm for}\,-\infty <x<-L-\frac{b}{2}\\
-\frac{1}{2p_0}\left(\frac{1}{q_0}+x+L+\frac{b}{2}\right)\sin \left[p_0\left(x+\frac{L+b}{2}\right)\right]\,&{\rm for}\,-L-\frac{b}{2}<x<-\frac{b}{2}\\
\frac{1}{2\sqrt{V_0}}\biggl[\left(\frac{p_0}{q_0}\left(x+\frac{L+b}{2}\right)+\frac{p_0}{q_0^2}-\frac{q_0 L}{2p_0}\right)e^{-q_0\left(x+\frac{b}{2}\right)}-\frac{V_0\left(1+q_0\frac{L}{2}\right)}{p_0 q_0^2}
e^{q_0\bigl(x+\frac{b}{2}\bigr)}\biggr]\,&{\rm for}\,\quad -\frac{b}{2}<x<\infty
\end{cases}\,,\\
&\partial_E \bphi^{(2)}_0(x)=-\partial_E \bphi^{(1)}_0(x)\,,
\end{align}
\end{subequations}
\end{widetext}
and
\begin{equation}
\partial_E\cN_0=
-\cN_0^3\,\frac{p_0^2-\left[q_0^2-p_0^2\left(1-q_0 b\right)\right]\left(1+q_0\frac{L}{2}\right)}{4p_0^2q_0^3}\,.
\label{dn}
\end{equation}

In Fig.~\ref{fig4_tails}, we compare the left-well WF $\Psi_L(x)$ obtained from our analytical approximated results in Eq.~\eqref{wan_approx_A} with the orbital $\Phi_0^{(1)}(x)$ from Eqs.~\eqref{orbitals_N2} and the exact solution obtained from the Schr\"odinger equation, for different values of barrier width $b$. Looking at Fig.~\ref{fig4_tails}, we observe that the left-well orbital $\Phi_0^{(1)}(x)$ approximates very well the corresponding WF in the left well region, confirming the high accuracy of the TPA for the evaluation of the tunneling Hamiltonian parameters. However, we note that $\Phi_0^{(1)}(x)$ has a very different tail into the neighboring well compared with the exact WF's one. In particular, the latter changes its sign in the right well, in agreement with Eq.~\eqref{orth} and our general arguments. This effect has deep implications for the evaluation of tunneling transition amplitudes in the presence of interparticle interaction. 
Finally, comparing Fig.~\ref{fig4_tails}(a) with Fig.~\ref{fig4_tails}(b), we note that in the case of narrow barrier, see Fig.~\ref{fig4_tails}(b), the tunneling energy increases (and consequently the WF's tail becomes more evident), so that the neglected higher-order terms become more important, and the analytical approximated results deviate from the exact solution.

Overall, the analytical approximated solution demonstrates high accuracy in modeling the exact WF, both in the well's and in the tail's regions. The exact solution, derived from the Schr\"odinger equation, describes the behavior of the system, including the modification in the WF's tail and its behavior across the barrier. On the other hand, the analytical approximated approach provides a robust and efficient alternative. This analytical method works very well across a broad range of conditions, effectively capturing the essential physics with significantly reduced computational effort. Thus, while the exact approach is essential for detailed studies, the analytical approximation is a powerful tool for theoretical explorations, especially in scenarios where computational resources are limited.

\section{NONSTANDARD HUBBARD MODEL}
\label{sec:NSHM}
So far, we have considered one particle placed in a double-well potential, deriving the corresponding WFs and comparing the analytical approximated results with the exact results and the orbitals.

In this section, we consider two particles placed in the same double-well system, analyzing the implications of the interparticle interaction on the total Hamiltonian of the system. Specifically, we use the results obtained in Sec.~\ref{sec:TPA} to derive simple analytical approximated expressions for the nonstandard DT and PT terms, as well as the standard on-site two-particle interaction energy $U$, in order to establish their magnitude as a function of the system's parameters.

Let us consider two distinguishable particles, hereinafter referred to with the labels $(1)$ and $(2)$ (e.g., two fermions with opposite spin projections), occupying the symmetric double-well potential of Fig.~\ref{fig3_sdw}, interacting through a two-body {\em repulsive} potential $V(x-y)>0$, where $x,y$ are the spatial coordinates of the two particles. In the site basis, the corresponding tunneling Hamiltonian, see Eq.~\eqref{tunn_ham} and Ref.~\cite{PRB_nonstandard}, can be written as
\begin{equation}
\begin{aligned}
\hat{H}&=E_{0}\left(\hn_L^{(1)}+\hn_L^{(2)}+
\hn_R^{(1)}+\hn_R^{(2)}\right)+\\
&+\Omega_{0}\left(\ha_L^{\dagger(1)}\ha_R^{(1)}+\ha_L^{\dagger(2)}\ha_R^{(2)}+H.c.\right)
+\hat V\,,
\label{tunn_ham_2part}
\end{aligned}
\end{equation}
where $$\hn_L^{(1,2)}=\ha_L^{\dagger(1,2)}\ha^{(1,2)}_L\,, \quad \hn_R^{(1,2)}=\ha_R^{\dagger(1,2)}\ha^{(1,2)}_R$$ are the number operators for the left and right well, $E_0$ is the site energy and $\Omega_0$ is the single-particle tunneling coupling. The interaction potential $\hV$ in Eq.~\eqref{tunn_ham_2part} is a sum of terms of form $$\ha_{i'}^{\dagger(1)}\ha_{j'}^{\dagger(2)}V_{i'j',ij}\ha_i^{(1)}\ha_j^{(2)}\,.$$ In the tunneling Hamiltonian basis, its matrix elements are:
\begin{equation}
V_{i'j',ij}=\int \Psi_{i'}(x)\Psi_{j'}(y)V(x-y)\Psi_i(x) \Psi_j(y)\,dx\,dy\,,
\label{V_int_terms}
\end{equation}
where $\Psi_i(x)$ is the $i$th site WF, and $i=L,R$.

The Hubbard interaction potential in second quantization formalism can be written as
\begin{equation}
\begin{aligned}
\hV_H &\equiv U\left(\hn_L^{(1)}\hn_L^{(2)}+\hn_R^{(1)}\hn_R^{(2)}\right)+\\
&+\bU\left(\hn_L^{(1)}\hn_R^{(2)}+\hn_L^{(2)}\hn_R^{(1)}\right)\,,
\label{hub_terms}
\end{aligned}
\end{equation}
with
\begin{equation}
U=\int n_L(x)n_L(y)V(x-y)\,dx\,dy\,,
\label{hub_U}
\end{equation}
and
\begin{equation}
\bU=\int n_L(x)n_R(y)V(x-y)\,dx\,dy\,,
\label{hub_ext}
\end{equation}
being the standard and the extended Hubbard terms, representing on-site and nearest-neighbor interaction, respectively, and $$n_{L,R}(x)=\Psi_{L,R}^2(x)\,.$$ The second term in Eq.~\eqref{hub_terms}, proportional to $\bU$, is not present in the standard Hubbard model Hamiltonian, and gives rise to the \textit{extended} Hubbard model \cite{dutta:non_standard}. Similarly to the standard Hubbard term, the latter does not generate tunneling transitions, rather it gives an extra contribution to the total energy of the system. 

In contrast, nonstandard Hubbard terms generate tunneling transitions between left and right well. The DT and PT terms can be respectively written as
\begin{subequations}
\label{nhab}
\begin{align}
\label{nhaba}
&\hV_{DT}=\Omega_1\Bigl[\left(\hn_L^{(1)}+\hn_R^{(1)}\right)\left(\ha_L^{\dagger(2)}\ha_R^{(2)}
+H.c.\right)+\nonumber\\
&\quad\quad\quad\quad+\left(\hn_L^{(2)}+\hn_R^{(2)}\right)\left(\ha_L^{\dagger(1)}\ha_R^{(1)}+H.c.\right)\Bigr] \,,\\
&\hV_{PT}=\Omega_2\left[\ha_R^{\dagger(1)} \ha_R^{\dagger(2)} \ha_L^{(1)}\ha_L^{(2)}+\ha_R^{\dagger(1)} \ha_L^{\dagger(2)} \ha_L^{(1)}\ha_R^{(2)}+H.c.\right] \,,
\label{nhabb}
\end{align}
\end{subequations}
with corresponding transition amplitudes given respectively by:
\begin{subequations}
\label{nhab_ampl}
\begin{align}
&\Omega_1=\int \Psi_L^2(x)\Psi_L(y)V(x-y)\Psi_R(y)\,dx\,dy\,,
\label{nhaba_ampl}\\
&\Omega_2=\int \Psi_L(x)\Psi_L(y)V(x-y)
\Psi_R(x)\Psi_R(y)\,dx\,dy\,.
\label{nhabb_ampl}
\end{align}
\end{subequations}
The first nonstandard Hubbard term DT, given by Eq.~\eqref{nhaba}, consists of a single-particle hopping process (e.g., $\Psi_{LL}\to\Psi_{LR}$) between the two wells resulting from the interaction of the tunneling particle with the nontunneling one (also known as bond-charge interaction, see Ref.~\cite{luhmann:multiorb_dens_ind}). Its tunneling amplitude ($\Omega_1$) sums up with the free single-particle tunneling amplitude, given by the tunneling energy $\Omega_0$ in the Hamiltonian in Eq.~\eqref{tunn_ham_2part}, resulting in an interaction-dependent effective single-particle tunneling \cite{dutta:non_standard, jurgensen:observation_density_induced_tunnelling}, expressed as
\begin{equation}
\Omega_{0}\left(\ha_R^{\dagger(1)} \ha_L^{(1)}+H.c.\right)\rightarrow\hat{\Omega}_{eff}\left(\ha_R^{\dagger(1)} \ha_L^{(1)}+H.c.\right)\,,
\end{equation}
where
\begin{equation}
\hat{\Omega}_{eff}=\Omega_{0}+\Omega_1\left(\hn_L^{(2)}+\hn_R^{(2)}\right)
\label{omega_effective}
\end{equation}
represents the effective single-particle tunneling operator, explicitly dependent on the occupation numbers of the two wells, through the factor $\hn_L^{(2)}+\hn_R^{(2)}$. Particularly, in a double-well potential with constant total density, the DT amplitude in Eq.~\eqref{nhaba_ampl} can always be incorporated into the single-particle tunneling amplitude, see Refs.~\cite{dutta:non_standard, jurgensen:observation_density_induced_tunnelling}, yielding an effective tunneling $$\Omega_{eff}\equiv\Omega_0+\Omega_1 \,.$$ Looking at Eq.~\eqref{omega_effective}, it is then clear that the interaction strength can be tuned in such a way to modify $\Omega_1$, and in turn $\Omega_{eff}$.

The second nonstandard Hubbard term, given by Eq.~\eqref{nhabb}, describes a simultaneous two-particle PT coherent hopping process, that can be both direct (e.g., $\Psi_{LL}\to\Psi_{RR}$) or exchange (e.g., $\Psi_{LR}\to\Psi_{RL}$). This term, with amplitude $\Omega_2$, arises when the interaction matrix is connecting two states where both particles switch their site location, representing an additional nonstandard physical process not present in the standard Hubbard Hamiltonian. Note that both nonstandard Hubbard amplitudes in Eqs.~\eqref{nhab_ampl} contain the overlap between the WFs belonging to neighboring wells. For this reason, they strongly depend not only on the shape of the interacting potential $V(x-y)$, but also 
on the WFs' tails in the neighboring wells. Therefore, an accurate evaluation and an analytical estimate of the WFs' tails are crucial in the determination of the nonstandard Hubbard terms.

\section{ANALYTICAL APPROXIMATED RESULTS WITH CONTACT INTERACTION}
\label{sec:CONTACT_INT}
Let us consider two distinguishable particles interacting through a 
$\delta$-shaped (\textit{contact}) interaction potential \cite{vermeyen:exchange, yang:strongly}. In contrast with the approach taken in Ref.~\cite{PRB_nonstandard}, our focus here is on the analytical evaluation of all terms included in the Hamiltonian. In fact, by employing $\delta$-shaped interaction, we simplify the mathematical treatment of Eq.~\eqref{V_int_terms}, enabling a more precise and manageable analysis of the dynamic behavior of the system. At first, we consider a \textit{repulsive} $\delta$-shaped interaction potential between particles, defined as
\begin{equation}
V(x-y)=V_\delta \delta (x-y)\,,
\label{delta_potential}
\end{equation}
where $V_\delta>0$ is the interaction strength. From Eq.~\eqref{delta_potential}, we can exactly evaluate the interaction potential matrix elements in Eq.~\eqref{V_int_terms}. Specifically, we focus on the standard Hubbard term in Eq.~\eqref{hub_U} and the DT and PT terms in Eqs.~\eqref{nhab_ampl}, and we compare the exact results with the analytical approximated ones.

We start our analysis with the on-site standard Hubbard term $U$, given by Eq.~\eqref{hub_U}. Using Eqs.~\eqref{wan_approx}, and neglecting $\cO(\Omega_0^2)$ terms, we find
\begin{equation}
U=V_\delta \cN_0^4\int\limits_{-\infty}^0 \left [\bphi_0^{(1)}(x)\right]^4\,dx=\frac{3}{4}\cN_0^2 V_\delta \left(1-\eta \right)\,,
\label{hubb_analytic}
\end{equation}
where $\cN_0$ is the normalization factor, see Eq.~\eqref{norm_factor}, and
$$\eta=\frac{p_0^2}{3V_0\left(1+\frac{q_0L}{2}\right)}\,.$$

We then evaluate the DT amplitude $\Omega_1$, given by Eq.~\eqref{nhaba_ampl}. Considering the symmetry of the WFs, namely $\Psi_R(x)=\Psi_L(-x)$, combined with Eqs.~\eqref{wan_approx}, and neglecting $\cO(\Omega_0^2)$ terms, we find
\begin{equation}
\begin{aligned}
\Omega_1&=V_\delta\int\limits_{-\infty}^\infty \Psi_L(x)^3\Psi_R(x)\,dx=\\
&=\Omega_0 V_\delta\int \limits_{-\infty}^0
\left(\bPhi_0^{(1)}(x)\right)^3\partial_E\bPhi_0^{(1)}(x)\,dx\,,
\end{aligned}
\label{omega1_an}
\end{equation}
From Eqs.~\eqref{wan_approx_norm}, we can rewrite $\Omega_1$ as the sum of two components, namely $$\Omega_1=\Omega_1^{(1)}+\Omega_1^{(2)}\,,$$
given by
\begin{subequations}
\label{omega1_2part}
\begin{align}
\Omega_1^{(1)}&= \Omega_0 V_\delta \cN_0^4\int\limits_{-\infty}^0\left[\bphi_0^{(1)}(x)\right]^3\partial_E \bphi_0^{(1)}(x)\,dx\,,\\
\Omega_1^{(2)}&= \Omega_0 V_\delta \cN_0^3\left(\partial_E \cN_0 \right)\int\limits_{-\infty}^0 \left[\bphi_0^{(1)}(x)\right]^4\,dx=\nonumber\\
&=\Omega_0 U\frac{\partial_E\cN_0}{\cN_0}\,.
\end{align}
\end{subequations}
These two components come from the dimensionless reduced orbital and the energy dependent normalization factor, respectively, see Eq.~\eqref{modif_orbitals}. Since, for \textit{contact} interaction, the DT term $\Omega_1$ is proportional to $\Omega_0$, it effectively renormalizes the tunneling coupling, see Eq.~\eqref{omega_effective}, so that
\begin{equation}
\Omega_0\to \Omega_{eff}=\Omega_0+\Omega_1\equiv\Omega_0\left(1+gV_\delta\right)\,.
\label{renorm}
\end{equation}
Note that, according to Eq.~\eqref{renorm}, if the ratio $\Omega_1/\Omega_0$ is negative, the effective tunneling coupling $\Omega_{eff}$ can be reduced by the interaction strength $V_\delta$. In other words, the DT term can either increase or decrease the single-particle tunneling coupling $\Omega_0$, depending on the sign of the coefficient $g$. Specifically, in the case of repulsive interaction ($V_\delta > 0$) and a negative coefficient ($g<0$), the DT term decreases the magnitude of the effective single-particle tunneling coupling $\Omega_{eff}$. With a sufficiently large interaction strength $V_\delta$, the single-particle tunneling can eventually be completely suppressed.
\begin{figure*}[t]
\centering
{\includegraphics[width=17.2cm]{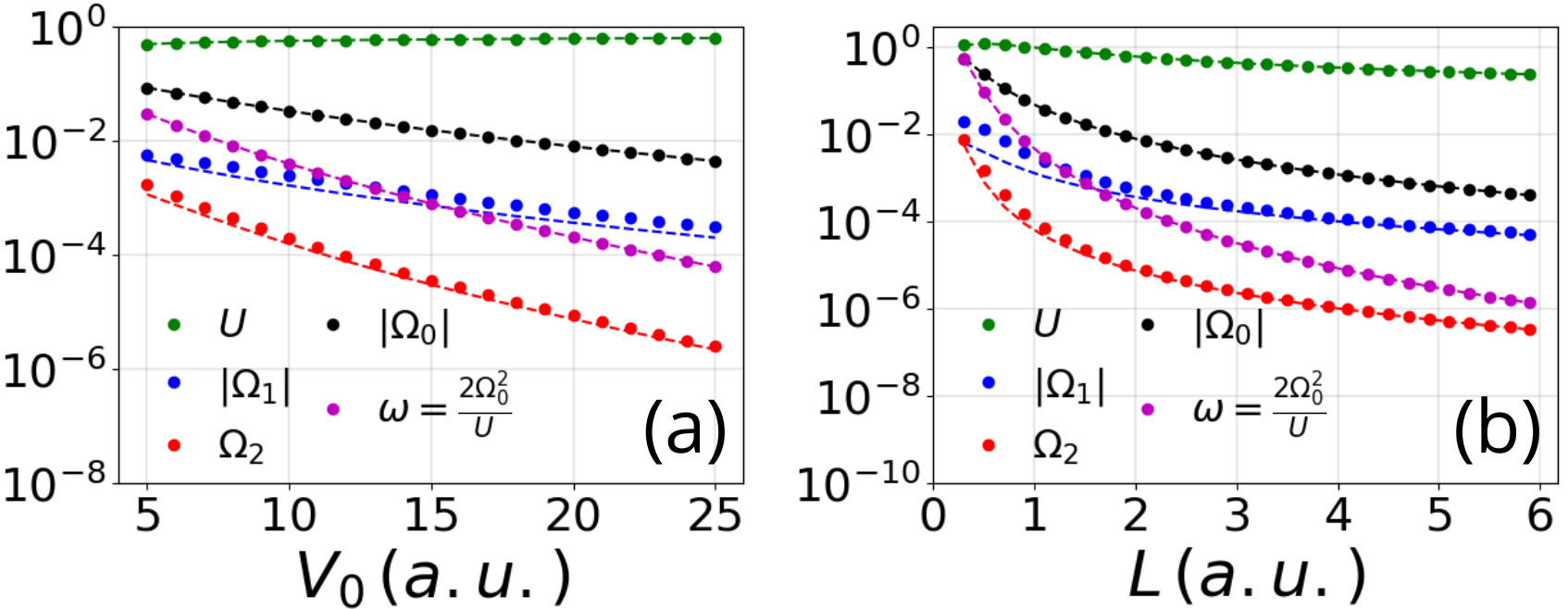}}
\caption{Comparison of exact results (dots) and analytical approximated results (dashed curves) for the Hubbard term $U$ (green curve), DT amplitude $|\Omega_1|$ (blue curve), PT amplitude $\Omega_2$ (red curve), single-particle tunneling $|\Omega_0|$ (black curve) and standard Hubbard second-order cotunneling $\omega$ (purple curve). The results are obtained considering two distinguishable particles placed in a square double-well potential, interacting via $\delta$-shaped interaction with $V_\delta=1$, as a function of the lattice depth $V_0$ (with $L=2$ and $b=1$) in panel (a) and of the well width $L$ (with $b=1$ and $V_0=20$) in panel (b).}
\label{fig5_comp_matrix}
\end{figure*}

Hence, the sign of the coefficient $g$ becomes crucial to understand how the DT term affects the dynamics of the system. At present, there is no general agreement on the sign of the coefficient $g$ (or even on its magnitude) \cite{hirsch:electron_hole, luhmann:multiorb_dens_ind, dutta:non_standard}. This problem has been investigated analytically in the framework of our simple toy-model, that allows a precise evaluation of the WFs' tails, in Ref.~\cite{PRB_nonstandard}. In that work, by analyzing Eq.~\eqref{orth} and Eq.~\eqref{omega1_an}, it was shown that $\Omega_1<0$ for contact interaction, so that the DT term can never match $\Omega_0$ and therefore suppress the effective single-particle tunneling $\Omega_{eff}$. On the other hand, only in presence of long-range interaction, it is possible to arrange the parameters of the well and the interparticle interaction in such a way to have $\Omega_{eff}=0$.

As said, the square double-well potential system allows us to obtain a simple analytical approximated expression for $\Omega_1$, which clearly shows the sign and the magnitude of the DT amplitude as a function of the system's parameters. Indeed, by using Eq.~\eqref{norm_factor}, Eqs.~\eqref{phi_der}, Eq.~\eqref{dn}, Eq.~\eqref{hubb_analytic} and Eq.~\eqref{omega1_an}, we find two analytical approximated expressions for $\Omega_1=\Omega_1^{(1)}+\Omega_1^{(2)}$, that read:
\begin{subequations}
\label{om1_approx_2}
\begin{align}
\Omega_1^{(1)}&=V_\delta \Omega_0 \cN_0^4\frac{\left(4p_0^4-3q_0^2V_0\right)\left(1+q_0\frac{L}{2}\right)+p_0^2\left(p_0^2+V_0\right)
}{32 p_0^2q_0^3V_0}=\nonumber\\
&=-\frac{3V_\delta\Omega_0 \cN_0^2}{32 p_0^2} \left[1+\cO\left(\frac{p_0^2}{V_0}\right)\right]\,,\\
\Omega_1^{(2)}&=-3V_\delta \Omega_0 \cN_0^2 \frac{\cN_0^2p_0^2-\left[q_0^3-p_0^2q_0\left(1-2q_0b\right)\right]}{16p_0^2q_0^3}(1-\eta)=\nonumber\\
&=\frac{6V_\delta\Omega_0 \cN_0^2}{32 p_0^2} \left[1+\cO\left(\frac{p_0^2}{V_0}\right)\right]\,.
\end{align}
\end{subequations}
In Eqs.~\eqref{om1_approx_2}, we neglect $\cO(p_0^2/V_0)$ terms, since we deal with the SB approximation, which considers only the lowest-band, where $p_0\ll \sqrt{V_0}$. Note that the previous condition means $|E_0|\simeq V_0$, which implies $q_0\simeq \sqrt{V_0}$. The expansion holds assuming also the condition $1+q_0L \gg 1$, corresponding to the Bohr quantization rule. Looking at Eqs.~\eqref{om1_approx_2}, we find that the first component $\Omega_1^{(1)}$ of the DT term, coming from the energy-dependence of the orbital, is always positive, while the second component $\Omega_1^{(2)}$, arising from a variation of the normalization factor with energy, is twice larger and of opposite sign ($\Omega_1^{(2)}\simeq -2\Omega_1^{(1)}$). As a result, the coefficient $g$ is always positive for contact interaction. This clearly shows the necessity of an accurate evaluation of the WFs' tails to determine the sign of DT term, which is hard to determine from general arguments, since the energy dependence of the normalization factors plays a crucial role.

Finally, we analyze the PT amplitude $\Omega_2$, given by Eq.~\eqref{nhabb_ampl}. Cotunneling refers to a general phenomenon that arises from second-order terms in the standard Hubbard model, as well as from the nonstandard PT term in the exact model. By applying Eqs.~\eqref{wan_approx_norm} and exploiting the symmetry property $\Psi_R(x) = \Psi_L(-x)$ of the WFs, we can write
\begin{equation}
\begin{aligned}
\label{om2_approx_2}
&\Omega_2=V_\delta\int\limits_{-\infty}^{+\infty}\Psi_L^2(x)\Psi_R^2(x)\,dx=2V_\delta \Omega_0^2 \cN_0^2\cdot\\
&\cdot\int\limits_{-\infty}^0\left(\bphi_0^{(1)}(x)\right)^2\left[\cN_0\partial_E\bphi_0^{(1)}(x)+\bphi_0^{(1)}(x)(\partial_E \cN_0)\right]^2\,dx=\\
&=\Omega_2^{(1)}+2\Omega_0\frac{\partial_E \cN_0}{\cN_0}
\left[2\Omega_1^{(1)}+\Omega_1^{(2)}\right],
\end{aligned}
\end{equation}
where
\begin{equation}
\Omega_2^{(1)}=2V_\delta \Omega_0^2 \cN_0^4\int\limits_{-\infty}^0 \left[\bphi_0^{(1)}(x)\right]^2\left[\partial_E\bphi_0^{(1)}(x)\right]^2\,dx\,.
\label{om2_approx_1}
\end{equation}
Since $\Omega_1^{(1,2)}$ are the two components of the DT term obtained in Eqs.~\eqref{om1_approx_2}, we only need to evaluate $\Omega_2^{(1)}$ in Eq.~\eqref{om2_approx_1}. By substituting Eqs.~\eqref{wan_approx}, Eqs.~\eqref{phi_der} and Eq.~\eqref{dn} into Eq.~\eqref{om2_approx_1}, the calculation can be performed analytically. By neglecting $\cO(\Omega_0^3)$ terms, we find
\begin{equation}
\Omega_2^{(1)}=\frac{\cN_0^4V_\delta\Omega_0^2}{2p_0^2q_0^3}\left[\frac{{\cal A}_0+{\cal A}_1\left(q_0\frac{L}{2}\right)}{32p_0^2V_0 q_0^2}+\frac{{\cal A}_2\left(q_0\frac{L}{2}\right)^2}{4V_0 q_0^2}+\frac{\left(q_0\frac{L}{2}\right)^3}{3}\right],
\end{equation}
where
\begin{equation}
\begin{split}
& {\cal A}_0=43p_0^6-p_0^4V_0(65-16q_0b)+13p_0^2V_0^2-V_0^3\,,\nonumber\\
& {\cal A}_1=92p_0^6-p_0^4V_0(161-32q_0b)+34p_0^2V_0^2-V_0^3\,,\nonumber\\
& {\cal A}_2=6p_0^4-p_0^2V_0(13-2q_0 b)+4V_0^2\,.\nonumber
\end{split}
\end{equation}
In the limit $|E_0| \ll V_0$, corresponding to the SB condition, the relation $2\Omega_1^{(1)}+\Omega_1^{(2)}\rightarrow 0$ holds. Therefore, Eq.~\eqref{om2_approx_2} becomes $\Omega_2=\Omega_2^{(1)}$, which is given by a simple analytical approximated expression: 
\begin{equation}
\Omega_2=\frac{2 \cN_0^4V_\delta\Omega_0^2}{4p_0^2 q_0^3}\left[\frac{\left(q_0\frac{L}{2}\right)^3}{3}+\frac{V_0\left(q_0\frac{L}{2}\right)^2}{q_0^2}-\frac{V_0^2\left(1+q_0\frac{L}{2}\right)}{32p_0^2 q_0^2}\right]\,.
\label{omega_2_approx_final}
\end{equation}
This result implies that the two-particle coherent tunneling amplitude $\Omega_2$ is proportional to the interaction strength $V_\delta$. Therefore, for large enough $V_\delta$, the PT amplitude $\Omega_2$ dominates over the corresponding cotunneling hopping amplitude of the standard Hubbard model, given by $$\omega\equiv \frac{2\Omega_0^2}{U}\,,$$ that decreases with the interaction strength $V_\delta$, as pointed out in the Introduction and since $U \propto V_\delta$, see Eq.~\eqref{hubb_analytic} and also Ref.~\cite{gurvitz:twoelectroncorrelated}. As expected, the DT term is $\propto \Omega_0$, see Eq.~\eqref{omega1_an}, while the PT term is $\propto \Omega_0^2$, see Eq.~\eqref{omega_2_approx_final}.

A comparison between exact and analytical approximated results for these terms is presented in Fig.~\ref{fig5_comp_matrix}. Specifically, we evaluate the on-site standard Hubbard term $U$, the noninteracting single-particle tunneling $\Omega_0$, the DT term $\Omega_1$, the PT term $\Omega_2$ and the second-order cotunneling $\omega=2\Omega_0^2/U$ as a function of the lattice depth $V_0$ in Fig.~\ref{fig5_comp_matrix}(a) and of the well width $L$ in Fig.~\ref{fig5_comp_matrix}(b). The exact results for $U$, $\Omega_0$, $\Omega_1$ and $\Omega_2$ are given by Eq.~\eqref{hub_U}, Eq.~\eqref{Wannier_functions_B}, Eq.~\eqref{nhaba_ampl} and Eq.~\eqref{nhabb_ampl}, respectively. In the same way, the analytical approximated results are given by Eq.~\eqref{hubb_analytic}, Eq.~\eqref{omega_0_bardeen}, Eq.~\eqref{om1_approx_2} and Eq.~\eqref{om2_approx_2}, respectively. Looking at Fig.~\ref{fig5_comp_matrix}, our analytical approximated results show an excellent match with numerical simulations across various parameters, including the lattice depth $V_0$ and the well width $L$. This strong agreement between the analytical approximated and numerical results confirms the validity and robustness of our method. In fact, it demonstrates that our analytical approach is capable of accurately capturing the essential physics of the system, providing a reliable alternative to computationally intensive simulations. However, we observe that this agreement fails for low values of $V_0$ and $L$. Specifically, for such low values of these parameters, the tunneling energy $\Omega_0$ increases. A low lattice depth $V_0$ results in a higher $\Omega_0$, while a low well width $L$ leads to higher-energy levels and, consequently, a larger tunneling amplitude, see Eq.~\eqref{Wannier_functions_B}. For these high values of $\Omega_0$, our analytical approximated approach fails, as the perturbation parameter is proportional to $\Omega_0$ itself, and it cannot be used as an expansion parameter.

\section{COMPARISON WITH EXISTING LITERATURE}
\label{sec:COMP_THEOR}
In this section, we extend our numerical results to different shapes of double-well potential, demonstrating the versatility of our numerical method. This expanded analysis shows the robustness of our approach across different configurations, underscoring its adaptability and the broader applicability of our findings beyond the initial setup. To do so, we compare our theoretical predictions with the results obtained in Ref.~\cite{luhmann:multiorb_dens_ind}, where the authors focus on the DT process of bosons in optical lattices and on how their density affects this process. Specifically, we adapt the shape of the double-well potential, considering a one-dimensional (1D) double-well potential defined as
\begin{equation}
\cV(x)=V_0 \,\cos^2\left(\frac{\pi x}{\lambda}\right)\,,
\label{cos2_pot}
\end{equation}
where $V_0$ is the lattice depth and $\lambda$ is the periodicity of the lattice. We perform our numerical simulations by considering two bosons interacting via $\delta$-shaped interaction. Note that our exact results obtained with distinguishable particles can be straightforwardly generalized for bosonic particles (as well as for fermionic particles, see Ref.~\cite{PRB_nonstandard}). It is important to highlight that our approach differs significantly in the choice of WFs used. While Ref.~\cite{luhmann:multiorb_dens_ind} employs the MLWFs, we adopt our new method for generating WFs.

Specifically, in Ref.~\cite{luhmann:multiorb_dens_ind}, the authors claim that the DT process can be relevant for the system's dynamics, since it can modify the free single-particle tunneling. In general, they demonstrate how nonstandard Hubbard models can describe more precisely the physics of bosonic atoms in lattice systems, including nonstandard terms not accounted for in the Hubbard model. In this work, the full lowest-band interaction Hamiltonian is defined as
\begin{equation}
\hat{H}_{int}=\frac{1}{2}\sum_{ijkl}V_{ijkl}\hat{b}_i^{\dagger}\hat{b}_j^{\dagger}\hat{b}_k\hat{b}_l\,,
\label{int_ham_exp}
\end{equation}
where $\hat{b}_i^{(\dagger)}$ is the annihilation (creation) operator for a bosonic particle in the ground state of the $i$-th site and
\begin{equation}
V_{ijkl}=8\pi a_s \int \Psi_i(x)\Psi_j(x)\Psi_k(x)\Psi_l(y)\,d^3x
\label{V_int_theor}
\end{equation}
are the interaction matrix elements, where $a_s$ is the free space $s$-wave scattering length, $m$ is the mass of the bosonic particle and $\Psi_i(x)$ is the lowest-band $i$-th site WF. Note that Eq.~\eqref{V_int_theor} can be traced back to Eq.~\eqref{V_int_terms}, considering that the interaction strength $V_\delta$ between two particles interacting via s-wave scattering is given by $V_\delta=8\pi a_s$ (see Ref.~\cite{esslinger:fermi_hubbard}). Eq.~\eqref{int_ham_exp} introduces also an off-site interaction between neighboring sites, which leads to different physical processes compared with the standard Hubbard model. Specifically, we focus on the DT and PT terms.

\begin{center}
\begin{figure}[t]
\centering
\includegraphics[width=8.6cm]{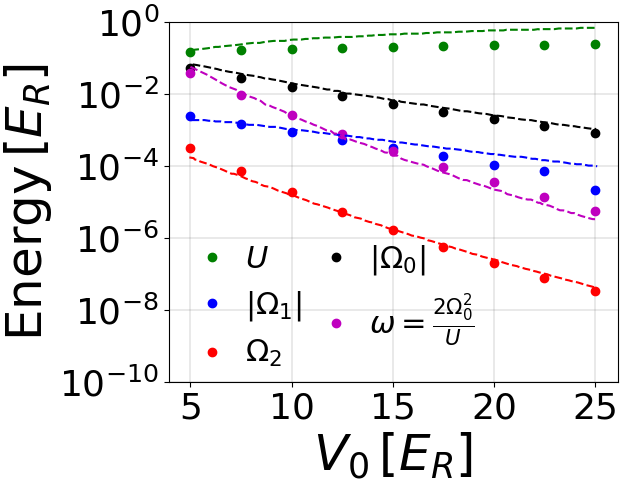}
\caption{Comparison of exact results obtained with the SB nonstandard model (dots) and results from Ref.~\cite{luhmann:multiorb_dens_ind} (dashed curves) for the Hubbard term $U$ (green curve), DT amplitude $|\Omega_1|$ (blue curve), PT amplitude $\Omega_2$ (red curve), single-particle tunneling $|\Omega_0|$ (black curve) and standard Hubbard model second-order cotunneling $\omega=2\Omega_0^2/U$ (purple curve). The results are obtained considering two bosons placed in the double-well potential of Eq.~\eqref{cos2_pot}, interacting via $\delta$-shaped interaction with $V_\delta=1$, as a function of the lattice depth $V_0$.}
\label{fig6_njp}
\end{figure}
\end{center}

In Fig.~\ref{fig6_njp}, we compare results presented in Ref.~\cite{luhmann:multiorb_dens_ind} with our model predictions. Specifically, we analyze the on-site standard Hubbard term $U$, the noninteracting single-particle tunneling $\Omega_0$, the DT term $\Omega_1$, the PT term $\Omega_2$ and the second-order cotunneling $\omega=2\Omega_0^2/U$ as a function of the lattice depth $V_0$. All the quantities are expressed in units of the recoil energy
$$E_r=\frac{h^2}{2m\lambda^2}\,,$$
where $m$ is the mass of the atoms, $h$ is the Planck's constant and $\lambda=765 \ nm$ represents the periodicity of the lattice.
Our results highlight the effectiveness of our numerical method in accurately capturing the essential dynamics of bosonic systems in optical lattices. The analytical predictions from our method show a remarkable agreement with the theoretical results found in the literature, particularly those presented in Ref.~\cite{luhmann:multiorb_dens_ind}. This demonstrates not only the validity but also the versatility and robustness of our approach. By using our novel method for generating WFs, we are able to handle various potential shapes and lattice configurations with high precision. This flexibility makes our method a powerful tool for studying complex lattice systems and their interactions, offering a reliable alternative to more traditional approaches.

\section{DYNAMICS OF TWO DISTINGUISHABLE PARTICLES IN A SQUARE DOUBLE-WELL POTENTIAL}
\label{sec:DYNAMICS}
As discussed in the previous sections, the presence of nonstandard Hubbard terms can introduce novel effects in the dynamics of a many-body system, totally nonaccounted for in the standard Hubbard model description. To analyze these effects further, we consider here the influence of the nonstandard Hubbard terms on the dynamics of two distinguishable particles in the symmetric square double-well potential shown in Fig.~\ref{fig3_sdw}. As a figure of merit of the system's dynamics, we consider the oscillation frequency of the time-dependent probability $P_{LL}(t)$, defined as the occupation probability to find both particles in the left well at time $t$. To evaluate it, let us consider the Hamiltonian given in Eq.~\eqref{tunn_ham_2part}, and the $\delta$-shaped two-particle interaction given in Eq.~\eqref{delta_potential}. The two-particle wave function, at time $t$, can be written as
\begin{equation}
\begin{aligned}
\ket{\Psi(t)}&=\Bigl[b_{LL}(t)\hat{a}_L^{\dagger(1)}\hat{a}_L^{\dagger(2)}+b_{LR}(t)\hat{a}_L^{\dagger(1)}\hat{a}_R^{\dagger(2)}\\
&+b_{RL}(t)\hat{a}_R^{\dagger(1)}\hat{a}_L^{\dagger(2)}+b_{RR}(t)\hat{a}_R^{\dagger(1)}\hat{a}_R^{\dagger(2)}\Bigr]\ket{0}\,,
\label{part_wf}
\end{aligned}
\end{equation}
where $\hat{a}_L^{(1)}\,,\hat{a}_R^{(1)}\,,\hat{a}_L^{(2)}$ and $\hat{a}_R^{(2)}\,(\hat{a}_L^{\dagger(1)}\,,\hat{a}_R^{\dagger(1)}\,,\hat{a}_L^{\dagger(2)}$ and $\hat{a}_R^{\dagger(2)})$ are the annihilation (creation) operators for the first/second particle in the left/right well, respectively, and $\ket{0}$ is the single-well vacuum state. The time-evolution of the two-particle wave function in Eq.~\eqref{part_wf} is given by the Schr\"odinger equation $$i\partial_t\ket{\Psi(t)}=\hH\ket{\Psi(t)}\,.$$
Considering $b_{LR}(t)=b_{RL}(t)$ (due to the potential symmetry), we can derive the equations of motion for the so called \textit{SB nonstandard model}, which include all the nonstandard interaction terms, which read
\begin{equation}
\begin{cases}
i \dot{b}_{LL}(t)= \left(2E_0 +U\right)b_{LL}(t) + \Omega_{eff} b_{LR}(t) + \Omega_2b_{RR}(t)\\
i \dot{b}_{RR}(t)= \left(2E_0 +U\right)b_{RR}(t) + \Omega_{eff}b_{LR}(t) + \Omega_2b_{LL}(t)\\
i \dot{b}_{LR}(t)= 2E_0 b_{LR}(t) + \Omega_{eff}\left[b_{LL}(t) + b_{RR}(t)\right]
\end{cases}\,,
\label{eq_mot_complete}
\end{equation}
where $E_0$ is the single-well ground-state energy, $U$ is the on-site standard Hubbard term given by Eq.~\eqref{hub_U}, $$\Omega_{eff}=\Omega_0+\Omega_1$$ is the effective tunneling coupling given by Eq.~\eqref{renorm} and $\Omega_2$ is the PT amplitude given by Eq.~\eqref{nhabb_ampl}. The equations of motion for the standard Hubbard model can be straightforwardly obtained from Eqs.~\eqref{eq_mot_complete} by simply setting $\Omega_1=\Omega_2=0$:
\begin{equation}
\begin{cases}
i \dot{b}_{LL}(t)= \left(2E_0 +U\right)b_{LL}(t) + \Omega_0 b_{LR}(t)\\
i \dot{b}_{RR}(t)= \left(2E_0 +U\right)b_{RR}(t) + \Omega_0 b_{LR}(t)\\
i \dot{b}_{LR}(t)=2E_0 b_{LR}(t) + \Omega_0\left[b_{LL}(t) + b_{RR}(t)\right]
\end{cases}\,.
\label{hubbard_eq_mot}
\end{equation}
\begin{figure*}[t]
\centering
{\includegraphics[width=17.2cm]{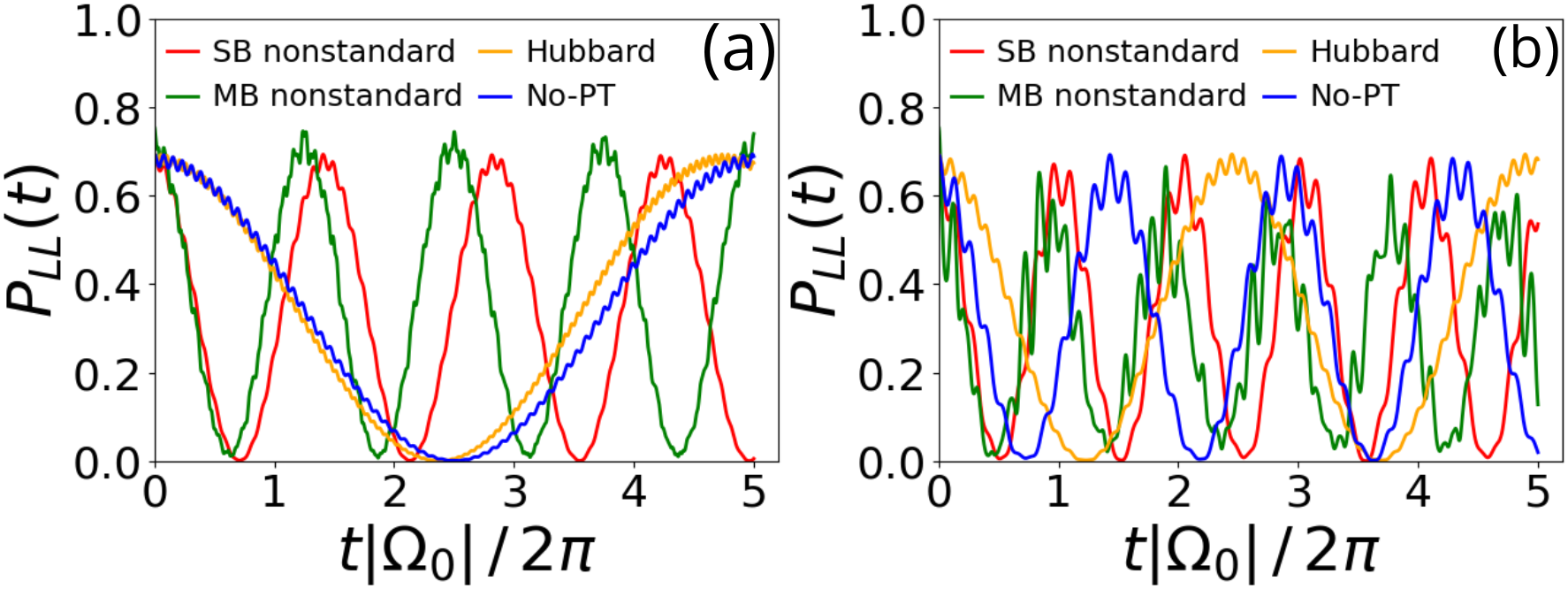}}
\caption{Time-evolution of the probability $P_{LL}(t)$ of finding at time $t$ two distinguishable particles together in the left well of the symmetric square double-well potential of Fig.~\ref{fig3_sdw}. The particles are interacting via $\delta$-shaped interaction potential. The different models are SB nonstandard model (red curve), MB nonstandard model (green curve), no-PT model (blue curve) and Hubbard model (orange curve). The double-well parameters are $L=2$, $b=0.5$ and $V_0=5$. The attractive interaction strengths are $U/|\Omega_0|\simeq -6$ (weakly interacting regime) in panel (a) and $U/|\Omega_0|\simeq -12$ (strongly interacting regime) in panel (b).}
\label{fig7_dynamics}
\end{figure*}

As initial state for the dynamics, we consider both particles placed in the infinite left-well state
$$\langle x \ket{\Phi}_L=\sqrt{\frac{2}{L}}\sin\left[\frac{\pi \left(x+L+b/2\right)}{L}\right]\,.$$
Therefore, the initial conditions for the equations of motion of Eqs.~\eqref{eq_mot_complete} and Eqs.~\eqref{hubbard_eq_mot} are $b_{LL}(0)=1$ and $b_{LR}(0)=b_{RL}(0)=b_{RR}(0)=0$. Finally, we can compute the time-dependent occupation probability $P_{LL}(t)$ for the state $$\ket{LL} \equiv \hat{a}_L^{\dagger(1)}\hat{a}_L^{\dagger(2)}\ket{00}\,.$$The system's parameters have been chosen in such a way to have four bound energies in the double-well system (for further details, see Appendix \ref{appendix_C}). 

In Fig.~\ref{fig7_dynamics}, we show the probability $P_{LL}(t)$ to find the two particles in the left well, for two different attractive interaction strengths, respectively in the weakly interacting regime ( Fig.~\ref{fig7_dynamics}a) and in the strongly interacting regime (Fig.~\ref{fig7_dynamics}b). Similar results can be obtained for the case of repulsive interaction strength and are reported in Appendix \ref{appendix_C}). Specifically, we compute the occupation probability $P_{LL}(t)$ by employing four different models: the SB nonstandard model, the MB nonstandard model, the \textit{no-PT} model and the standard Hubbard model. While the standard Hubbard model serves as a reference, the other three models include novel effects given by the exact treatment of the interaction. In particular, the SB nonstandard model considers only the effect of the lowest energy band, neglecting contributions from higher-energy bands. On the other hand, the MB nonstandard model fully accounts for the influence of all energy bands, see Ref.~\cite{PRB_nonstandard}. Moreover, we introduce the \textit{no-PT} model, to highlight the importance of the PT process itself, by artificially excluding its contribution from the dynamics of the SB nonstandard model.

We note that, for both weakly and strongly interacting regimes, the expected frequency of oscillation of $P_{LL}(t)$ from the standard Hubbard model (yellow curve) is smaller than the one obtained from the SB and MB nonstandard models (respectively red and green curves). This highlights the role of the exact treatment of interaction in the system's dynamics, justifying the inclusion of the nonstandard terms in the system Hamiltonian. Specifically, we note that for weakly interacting regime, the frequency is closer to the SB nonstandard model one, being the nonstandard term less relevant in the dynamics, compared with the strongly interacting regime. We also note that, for the strongly interacting regime, see Fig.~\ref{fig7_dynamics}(b), the no-PT model matches quite well the Hubbard model predictions. This puts in evidence the role (and the importance) of PT in the strongly interacting regime. Finally, we appreciate the difference between SB and MB nonstandard models, given by the different number of energy levels included in the dynamics, especially in the strongly interacting regime (the excitation probability for the higher levels is increased).

\begin{figure*}[t]
\centering
\includegraphics[width=17.2cm]{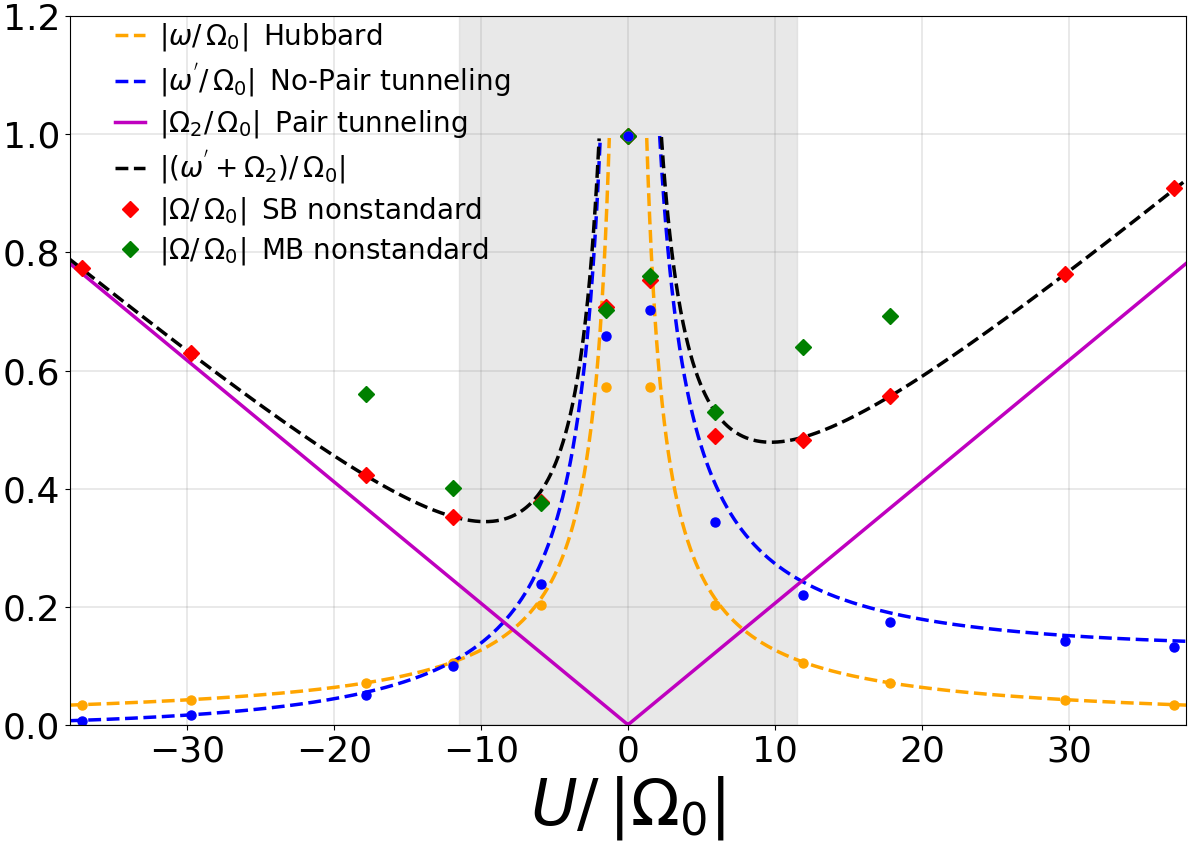}
\caption{Frequencies of the different processes involved in the dynamics of the system. Standard Hubbard model cotunneling $\omega=2\Omega_0^2/|U|$ (orange dashed curve), no-PT model $\omega'=2\Omega_{eff}^2/|U|$ (blue dashed curve), PT $\Omega_2$ (purple curve). Orange and blue dots stand for the Hubbard model and no-PT model, respectively, extracted from our simulations. Frequency of the SB nonstandard model for $P_{LL}(t)$, computed via Fourier transform (red diamonds). Frequency of the MB nonstandard model (four energy levels in total) for $P_{LL}(t)$, computed via Fourier transform (green diamonds). Black dashed curve: $\Omega=\omega'+\Omega_2$. The parameters of the double-well system are: $L=2$, $b=0.5$, $V_0=5$. With these parameters, $\Omega_0\simeq 0.22$, $|\Omega_1/U|\simeq 0.017$, $|\Omega_2/U|\simeq 0.02$ (DT and PT result both relevant in the dynamics).}
\label{fig8_main_fig}
\end{figure*}

To understand how the different tunneling processes affect the global system's dynamics, and how the two frequencies $\Omega_{eff}$ and $\Omega_2$ (interaction-dependent) act on the dynamics, we extracted the dominant frequency of the probability $P_{LL}(t)$ for different values of interaction strengths, both attractive and repulsive. This has been obtained by taking the frequency corresponding to the largest amplitude in the Fourier spectrum of $P_{LL}(t)$. In Fig.~\ref{fig8_main_fig}, we represent the dominant frequency of $P_{LL}(t)$ as a function of the interaction strength $U/|\Omega_0|$.
Looking at Fig.~\ref{fig8_main_fig}, we clearly distinguish three different regimes: for a strongly enough interacting regime $|U| \gg \Omega_0$, both attractive and repulsive, the dynamics is completely dominated by the PT process (purple line). Indeed, while the frequency $\omega={2\Omega_0^2}/{U}$ predicted by the standard Hubbard model (orange curve) decreases linearly with the on-site standard Hubbard interaction $U$ (and consequently with $V_\delta$), the frequency $\Omega_2$ increases linearly with it. For this reason, in the strongly interacting regime the PT process becomes predominant. By contrast, in the weakly interacting regime $|U| \ll \Omega_0$, the SB nonstandard (and MB nonstandard) model dynamics is very well approximated by the standard Hubbard model, which takes into account only single-particle tunneling processes. In this weakly interacting regime, the Hubbard model proves to be a good approximation for the system's dynamics.

Finally, we can distinguish an intermediate regime of interaction where both processes (cotunneling and PT) are equally involved in the dynamics. We note that the expected frequency of the SB nonstandard model corresponds to the sum of the two contributions: $\Omega=\omega' + \Omega_2$ (black dashed curve), where $$\omega'=\frac{2\left(\Omega_{eff}\right)^2}{|U|}\,.$$
Finally, the vertical gray band in Fig.~\ref{fig8_main_fig} corresponds to the case $|U|=\Delta E$, where the Hubbard term $U$ matches the mean level spacing $\Delta E$ of the double-well potential (taken as the distance in energy between the second-excited state and the average of the ground state and first-excited state of the double-well system). In this particular case, the interparticle interaction is strong enough to excite the higher-energy levels, making them play a relevant role in the dynamics of the system. Therefore, the SB approximation is no more valid, and the contribution given by the higher-energy levels must be included in the physical description through the MB nonstandard model. As clearly visible in Fig.~\ref{fig8_main_fig}, near the edges of this gray band, the results obtained with the SB nonstandard model start to deviate from the exact ones, obtained via the MB nonstandard model. Indeed, in this case the interaction between particles is sufficiently strong to excite the upper levels of the spectrum that cannot be neglected in the study of the dynamics. Finally, when $\delta$-shaped interaction is adopted in higher dimensions, regularization is necessary, and multiband effects must be included in the scattering process (see Refs.~\cite{Buchler:microscopic, haydn:microscopic}.)

\section{COMPARISON WITH EXPERIMENTS. TWO BOSONS IN A DOUBLE-WELL POTENTIAL}
\label{sec:COMP_EXP}
\begin{figure*}[t]
\centering
{\includegraphics[width=17.2cm]{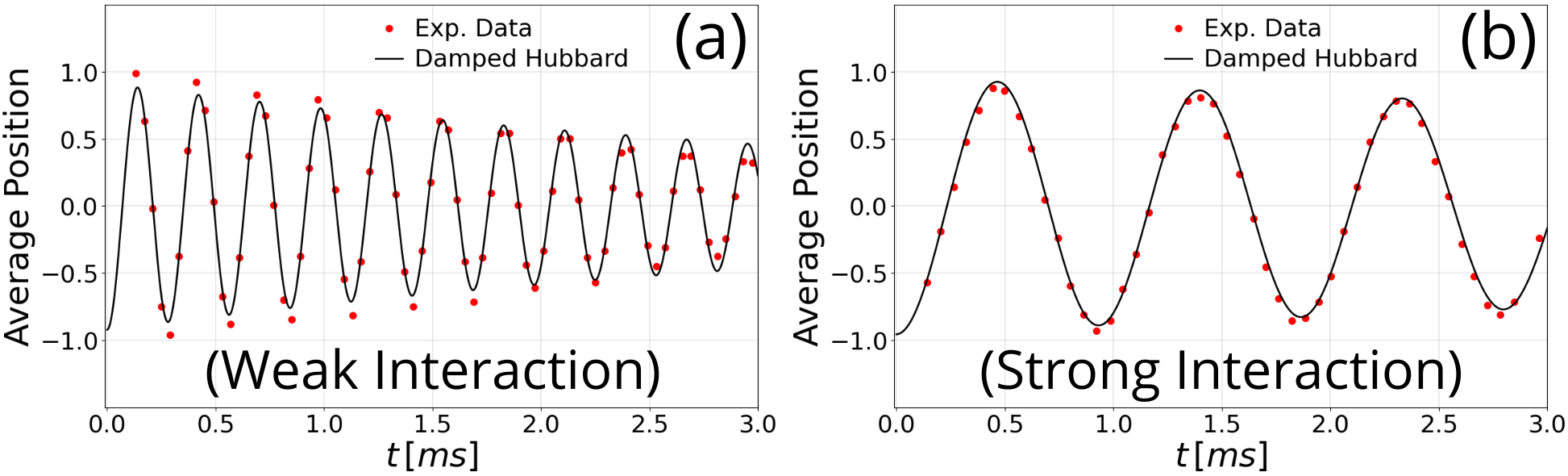}}
\caption{Comparison between experimental results of Ref.~\cite{bloch:direct_observation} (red dots), and damped Hubbard model (black curve) for a single boson initially localized in the left well. The parameters used are $A_{long}=9.5E_R$, $B_{short}=5.40E_R$, $\Delta \simeq 0.17E_R$ in panel (a) and $A_{long}=9.5E_R$, $B_{short}=7.92E_R$, $\Delta \simeq 0.07E_R$ in panel (b). All the energies are expressed in units of the recoil energy $E_R=h^2/2m\lambda^2$, where $h$ is the Planck's constant, $m\simeq 86.9u$ is the mass of the ${}^{87}$Rb atoms and $\lambda=765$ nm is the long-lattice wavelength. The decay times of the damped Hubbard model are: $\tau \simeq 4.38\,ms$ in panel (a) and $\tau \simeq 13\,ms$ in panel (b).}
\label{fig9_bloch_1part}
\end{figure*}
\begin{figure*}[t]
\centering
{\includegraphics[width=17.2cm]{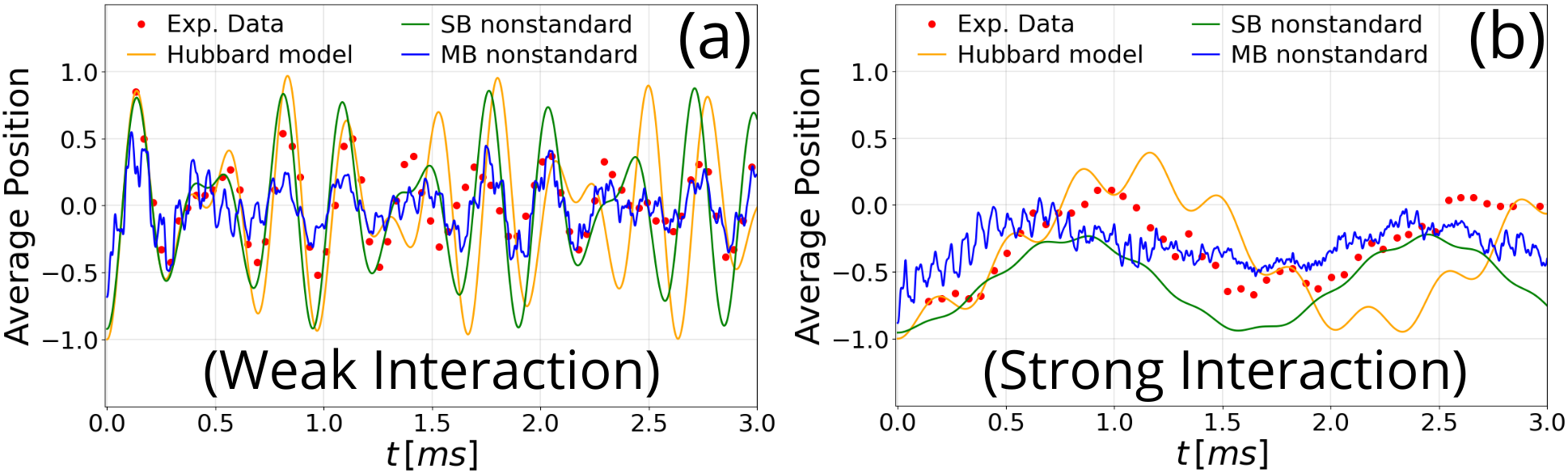}}
\caption{Comparison between experimental results of Ref.~\cite{bloch:direct_observation} (red dots), Hubbard model (orange curve), SB nonstandard model (green curve) and MB nonstandard model (blue curve) for two bosons in the weakly interacting regime ($U/J=0.67$ in panel (a)) and strongly interacting regime ($U/J=5$ in panel (b)), after initially preparing the system with both particles localized in the left well of the double-well system. The system's parameters are: $V_0 \simeq 22.4 E_R$, $\Omega_0 \simeq -0.45 E_R$, $\cE_1 \simeq -20.3 E_R$, $\cE_2 \simeq -19.4 E_R$ in panel (a) and $V_0 \simeq 26.8 E_R$, $\Omega_0 \simeq -0.14 E_R$, $\cE_1 \simeq -23.5 E_R$, $\cE_2 \simeq -23.2 E_R$ in panel (b).}
\label{fig10_bloch_2part}
\end{figure*}
In this section, our numerical results are compared with the experimental results reported in Ref.~\cite{bloch:direct_observation}. In that work, second-order atom tunneling processes were observed in an interacting ultracold bosonic gas of rubidium atoms placed in an optical double-well potential. It was observed that, under certain conditions, atoms undergo second-order tunneling processes in addition to the usual first-order tunneling expected from the standard Hubbard model. The authors identify a regime where single-particle tunneling is promoted by the presence of other particles (via the DT mechanism) and a PT process also affects the dynamics of the system.

To compare the experimental outcomes with the results obtained from our approach, we exploit its versatility and robustness, taking advantage of its ability to handle any potential shape, and not just square potentials. This flexibility is crucial for accurately modeling real systems, where, in general, the potential is given by a superposition of two or more optical potentials. From now on, if not stated otherwise, all the energies will be expressed in units of the recoil energy $E_R=h^2/2m\lambda^2$, where $\lambda$ represents the with of the double-well system. Here, our unit of length $a=\lambda/2=L+b$ represents the distance between two particles placed in neighboring lattice sites.

In Ref.~\cite{bloch:direct_observation}, the authors consider the Bose-Hubbard model as the most simple description of a set of $N$ bosonic atoms in an optical lattice with tunneling coupling $J\equiv \Omega_0$ between nearest-neighbor sites and two-body interaction $U$ between particles at the same site. The Hamiltonian that describes this model, considering a bias $\Delta$ between neighboring wells, is given by
\begin{equation}
\hat{H}_{BH}=-J\sum_{\braket{i,j}}\hat{b}^{\dagger}_i\hat{b}_j+\frac{U}{2}\sum_i\hat{n}_i\left(\hat{n}_i-1\right)-\frac{\Delta}{2}\sum_{\braket{i,j}}\left(\hat{n}_i-\hat{n}_j\right)\,.
\end{equation}
Restricting our considerations to a double-well potential, the bosonic atoms in a double-well potential are described in Ref.~\cite{bloch:direct_observation} by the standard Bose-Hubbard Hamiltonian
\begin{equation}
\begin{aligned}
\hat{H}_{BH}=&-J\left(\hat{b}_L^{\dagger}\hat{b}_R+\hat{b}_R^{\dagger}\hat{b}_L\right)-\frac{\Delta}{2}\left(\hat{n}_L-\hat{n}_R\right)\\
&+\frac{U}{2}\left[\hat{n}_L\left(\hat{n}_L-1\right)+\hat{n}_R\left(\hat{n}_R-1\right)\right]\,,
\end{aligned}
\end{equation}
where $J$ is the single-particle tunneling, $\hat{b}_{L,R}^{(\dagger)}$ are the annihilation (creation) operators for a bosonic particle in the ground state of the left and right well, $U$ is the interaction energy of two particles placed in the same well and $\Delta$ is the bias between the potential wells. The double-well potential in Ref.~\cite{bloch:direct_observation} is realized by superimposing two optical periodic potentials, respectively long-lattice and short-lattice periodic potentials, with amplitudes $A_{long}$ and $B_{short}$, and corresponding wavelengths given by $765.0$ nm and $382.5$ nm, respectively. The initial state for the dynamics is realized by placing one (or more) atoms in the left well. The sudden lowering of the potential barrier depth gives rise to the dynamics of the particle(s). As a figure of merit for the system's dynamics, the average position is considered:
\begin{equation}
\braket{x(t)}=\frac{\braket{\hat{n}_R(t)}-\braket{\hat{n}_L(t)}}{\braket{\hat{n}_R(t)}+\braket{\hat{n}_L(t)}}\,,
\end{equation}
where $\hat{n}_{L,R}$ is the number of bosons placed in the left (L) and right (R) well, respectively. Moreover, the dynamics of a single boson (single-particle signal) and of a pair of bosons (double-particle signal), for different interaction regimes, have been experimentally measured. Specifically, weakly ($U/J=0.67$) and strongly ($U/J=5$) interacting regimes have been considered. Note that, even in the single-particle case, the ratio $U/J$ can be defined, since the single-particle signal is obtained by subtracting the double-particle signal from the total signal (for further details, see Ref.~\cite{bloch:direct_observation}). One may wonder what occurs in the parameter range between the weakly and strongly interacting regimes considered in Fig.~\ref{fig10_bloch_2part}. According to the experimental conditions outlined in Ref.~\cite{bloch:direct_observation}, increasing the Hubbard interaction term $U$ also increases the lattice depth $V_0$, which in turn reduces the relative influence of the PT term. This issue is discussed in detail in Appendix \ref{appendix_D}, where it is also shown that the DT term becomes relevant in determining the parameter value at which the metal-insulator transition occurs.

In addition, our numerical model can be defined as a generalized lowest-band Hubbard Hamiltonian, which reads
\begin{equation}
\begin{aligned}
\hat{H}=&-J\left(\hat{b}_L^{\dagger}\hat{b}_R+\hat{b}_R^{\dagger}\hat{b}_L\right)-\frac{\Delta}{2}\left(\hat{n}_L-\hat{n}_R\right)\\
&+\frac{U}{2}\left[\hat{n}_L\left(\hat{n}_L-1\right)+\hat{n}_R\left(\hat{n}_R-1\right)\right]+\bU\left(\hat{n}_L\hat{n}_R\right)\\
&+\Omega_1\left(\hat{b}^{\dagger}_L\left(\hat{n}_L+\hat{n}_R\right)\hat{b}_R+\hat{b}^{\dagger}_R\left(\hat{n}_R+\hat{n}_L\right)\hat{b}_L\right)\\
&+\Omega_2\left(\hat{b}^{\dagger 2}_L\hat{b}^2_R+\hat{b}^{\dagger 2}_R\hat{b}^2_L\right)\,,
\end{aligned}
\end{equation}
where $\bU$ represents the nearest-neighbor interaction, $\Omega_1$ is the DT term and $\Omega_2$ is the PT term. For the case of single-particle dynamics, we modify both the amplitude $B_{short}$ of the short-lattice periodic potential and the bias $\Delta$ between the wells, in order to adjust the single-particle tunneling $\Omega_0$. We also introduce a phenomenological exponential damping in our model, to fit the experimental conditions. Specifically, we fit the bare model $M(t)$ with the function $$D(t)=M(t)e^{-\gamma t}\,,$$ where $\gamma$ represents the inverse decay time, and it is used as a free fitting parameter. The results for the dynamics of a single boson are presented in Fig.~\ref{fig9_bloch_1part}, both for weakly and strongly interacting regime. Here, the damped Hubbard model (black curve) fits very well the experimental data (red dots), matching the single-particle oscillation frequency. Specifically, in Fig.~\ref{fig9_bloch_1part}(a), we obtain a parameter $\gamma=0.2285\,ms^{-1}$, which corresponds to a decay time $\tau \simeq 4.38\,ms$ close to the decay time $\tau=3.5\,ms$ obtained in Ref.~\cite{bloch:direct_observation}. 

The average position of two interacting bosons, initially placed in the infinite left-well state, respectively in the case of weakly ($U/J=0.67$) and strongly ($U/J=5$) interacting regime, is shown in Fig.~\ref{fig10_bloch_2part}. Here, we represent the experimental data (red dots), the Hubbard model (orange curve), the SB nonstandard model (green curve) and the MB nonstandard model (blue curve). We note that the Hubbard model does not reproduce exactly the experimental results, both in terms of amplitude and frequency of oscillation, for weakly and strongly interacting regimes. Moreover, for the parameters chosen in the experiments, also the SB nonstandard model does not agree completely with the experimental results. On the contrary, the MB nonstandard model is able to fit better the experimental data, both in the weakly and strongly interacting regimes, since it considers all the possible energy levels involved in the system's dynamics. Comparing Fig.~\ref{fig10_bloch_2part}(a) with Fig.~\ref{fig10_bloch_2part}(b), we note that the SB nonstandard and MB nonstandard model approximate better the experimental data for the weakly interacting regime, with respect to the strongly interacting one. This highlights the importance of including higher-energy levels for the case of strongly interacting regime, to accurately capture the complex dynamics of strongly interacting bosonic systems. Even if our results do not exactly match the experimental data, they still highlight a significant improvement in the description of the system's dynamics compared with the standard Hubbard model.

\section{CONCLUSIONS}
In conclusion, our comprehensive study of many-body tunneling dynamics in arbitrary double-well potentials significantly advances the understanding of interacting many-body systems beyond the conventional approximations of the standard Hubbard model. By incorporating nonstandard Hubbard terms in the Hamiltonian, specifically density-induced tunneling and pair tunneling terms, we identify crucial modifications to the system's behavior that are not captured by the standard Hubbard model. Specifically, in the presence of a $\delta$-shaped repulsive interparticle interaction, our perturbative analytical approximations, corroborated by extensive numerical simulations, reveal that these additional terms fundamentally modify the single-particle tunneling parameter $\Omega_0$ and introduce new coherent propagation mechanisms, given by the pair tunneling process. These findings may have important implications for a broad range of physical phenomena, including high-$T_C$ superconductivity \cite{bistritzer:moire_bands, vu:moire_mott, chan:pairing, cao:uncon_superc, doping:lee, dagotto:highTc, bednorz:possible_high_Tc, cao:corr_insulator, anderson:resonating, brinkman:electronholecouplingcuprates, kaczmarczyk:2d_hubbard_gutzwiller, dagotto:high_tc_superconductors} and metal-insulator transitions \cite{zhou:pair_tunnelling_bosons, Jaksch:zoller_1998, anderson:valence, imada:review, petrosyan:liquid}.

We show that the nonstandard Hubbard model significantly deviates from the standard Hubbard model with increasing interaction strength, resulting in different transport behaviors. In the nonstandard model, strong interactions modify single-particle tunneling and enhancing pair tunneling, generating an interplay that may give rise to novel transport phenomena. However, at lower interaction strengths, the two models produce similar outcomes, particularly when density-induced tunneling and pair tunneling terms are minimal compared with $\Omega_0$.

Our theoretical framework has been also validated by experimental observations of second-order atom tunneling in optical double-well arrays, demonstrating the practical relevance and applicability of our model. The excellent agreement between our numerical simulations and the computed lowest-band parameters further underscores the robustness and accuracy of our approach. Overall, this work not only highlights the necessity of considering nonstandard Hubbard terms in the study of many-body systems, but also provides insights into the complex interplay between interaction and tunneling processes in quantum systems. These findings pave the way for future investigations into complex quantum behavior and emergent phenomena in controlled experimental setups.

From a broader perspective, the results of this study open alternative paths for exploring exotic quantum states and novel phases of matter in various physical systems. By challenging and extending the boundaries of the Hubbard model, our work suggests a reevaluation of theoretical frameworks that describe strongly correlated materials. This can potentially lead to the discovery of new materials with unique properties, thereby impacting the fields of condensed-matter physics. Additionally, the methods developed here could inspire advancements in quantum simulation and computation, where precise control and understanding of tunneling dynamics are crucial. As such, our findings not only improve the theoretical understanding of many-body quantum systems, but also have far-reaching implications for future experimental and technological developments.

\begin{acknowledgments}
F.B., M.Z. and G.L.C. acknowledge the support of the Iniziativa Specifica INFN-DynSysMath. This work has been financially supported by the Catholic University of Sacred Heart and by M.I.U.R. within Project No. PRIN 20172H2SC4. M.Z. acknowledges the Ermenegildo Zegna's Group for financial support. S.G. would like to thank Y. Oreg and E. Berg for helpful discussions and suggestions. We also thank S. Mailoud and G. Farinacci for discussions and for their valuable contribution at the initial stage of this work.
\end{acknowledgments}


%

\onecolumngrid
\newpage
\appendix

\section{Squared sine double-well potential}
\renewcommand{\theequation}{A.\arabic{equation}}
\setcounter{equation}{0}
\renewcommand{\thefigure}
{A.\arabic{figure}}
\setcounter{figure}{0}
\label{appendix_A}
In this section, we compare the results of our analytical approximated approach (TPA) with the exact numerical evaluation of the WFs $\Psi_j(x)$ for the following squared sine double-well potential, shown in Fig.~\ref{figA1_tails_cos2}(a):
\begin{equation}
\cV(x)=
\begin{cases}
-V_0\sin^2(x) \quad\quad & \text{for} \quad |x|\le \pi \\
0 & \text{for}\quad |x|> \pi
\end{cases}\,,
\label{V(x)_pot_cos2}
\end{equation}
where $V_0>0$ is the potential depth. The potential in Eq.~\eqref{V(x)_pot_cos2} can be seen as a sum of two single-well potentials, namely $\cV(x)=\cV_1(x)+\cV_2(x)$, where
\begin{equation}
\cV_1(x)=
\begin{cases}
-V_0\sin^2(x) \quad\quad & \text{for} \quad -\pi\le x\le 0\\
0 & \text{for}\quad x<-\pi \vee x>0
\end{cases}\,,
\label{V1(x)_cos2}
\end{equation}
while $\cV_2(x)=\cV_1(-x)$. To simplify the notation, in the following we use dimensionless units, i.e., $\hbar=2m=1$, unless otherwise specified. First, we consider the exact numerical solution for $E_\pm$ and the wave functions $\psi_\pm(E_\pm,x)$, given by Eqs.~\eqref{eigen_site} of the main text, where $\bPhi^{(1)}(E,x)$ is obtained from the numerical solution of the Schr\"odinger equation
\begin{equation}
\begin{aligned}
-&\bPhi^{(1)\prime\prime}(E,x)=\left(E-\left(V_0\cos^2(x)-V_0\right)\right)\bPhi^{(1)}(E,x) \quad {\rm for} \quad -\pi\le x\le 0\,,\\ 
&\bPhi^{(1)}(E,x)=e^{q (x+\pi)} \quad {\rm for}\quad 
x\le -\pi \,,
\end{aligned}
\end{equation}
where $q=\sqrt{-E}$. The eigenenergies are obtained from the matching conditions, namely $\bPhi^{(1)\prime}(E_+,0)=0$ for the symmetric state $\psi_+(E_+,x)$ and $\bPhi^{(1)}(E_-,0)=0$ for the antisymmetric state $\psi_-(E_-,x)$. Using Eqs.~\eqref{wan_approx} of the main text, we then obtain the exact WFs $\Psi_{L,R}(x)$. The eigenenergy levels $$E_\pm=\cE_{1,2}+\cO(\beta^2)$$ are denoted by dashed black lines in Fig.~\ref{figA1_tails_cos2}(a), while the exact left-well WF $\Psi_L(x)$ is displayed in Fig.~\ref{figA1_tails_cos2}(b) (red solid curve) for lattice depth $V_0=4.5$. The corresponding left-well orbital $\Phi_0^{(1)}(x)$, defined in Eqs.~\eqref{orbitals_N2} of the main text, is also shown (black dashed curve).
\begin{figure*}[h]
\centering
{\includegraphics[width=17.2cm]{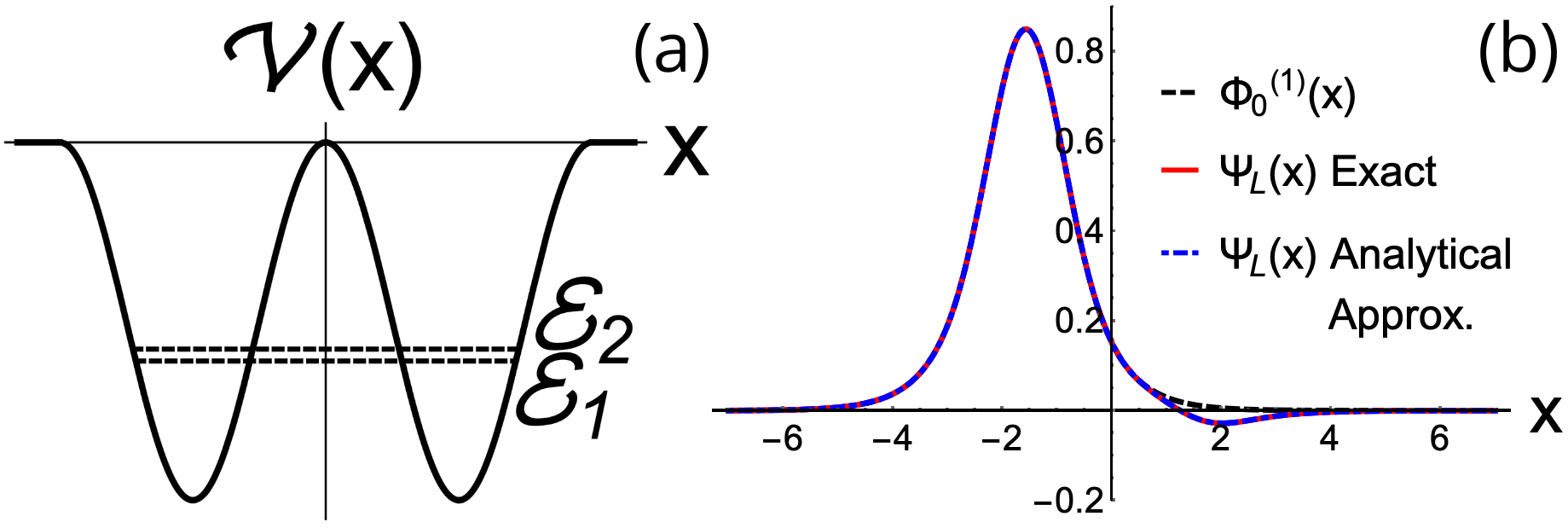}}
\caption{(a) Squared sinusoidal double-well potential with first two exact eigenenergies $\cE_{1,2}$ (black dashed lines). The bound states energies are: $\cE_1 \simeq -2.75$ and $\cE_2 \simeq -2.60$, respectively. (b) Comparison between the orbital $\Phi_0^{(1)}(x)$ (black dashed curve), the exact left-well WF $\Psi_L(x)$ (red solid curve) and the analytical approximated (neglecting $\cO(\Omega_0^2)$ terms) left-well WF (blue dot-dashed curve) for lattice depth $V_0=4.5$. The exact ground-state energy and tunneling coupling are $\bE_0 \simeq -2.672$ and $\bom_0 \simeq -0.075$, respectively, while the corresponding parameters given by the TPA are $E_0 \simeq -2.672$ and $\Omega_0 \simeq -0.075$.}
\label{figA1_tails_cos2}
\end{figure*}
We note that the WF almost corresponds to the respective orbital inside the respective well, while its tail in the other well differs from the orbital's one, as expected from our previous analysis. To compare these results with our approximation in Eqs.~\eqref{wan_approx} of the main text, we consider the tunneling Hamiltonian in Eq.~\eqref{tunn_ham_2part} of the main text, whose parameters $E_0$ and $\Omega_0$ are directly related to $E_\pm$ via Eq.~\eqref{Wannier_functions_A} and Eq.~\eqref{Wannier_functions_B} of the main text, see also Eqs.~\eqref{params_Ham} of the main text. Using $\Omega_0$ as energy shift in the orbital, we can evaluate the corresponding WF by using Eqs.~\eqref{wan_approx}, or equivalently Eqs.~\eqref{wan_approx_norm} of the main text. The resulting analytical approximated WF is represented in Fig.~\ref{figA1_tails_cos2}(b) (blue dot-dashed curve). For this set of parameters, our analytical approximated approach produces a result which is very close to the exact one. We expect our approach to be the more accurate the more the barrier height ensures a good localization of the left and right states in the respective well.

\section{Sinusoidal double-well potential}
\renewcommand{\theequation}{B.\arabic{equation}}
\setcounter{equation}{0}
\renewcommand{\thefigure}
{B.\arabic{figure}}
\setcounter{figure}{0}
\label{appendix_B}
In this section, we compare the results of our analytical approximated approach (TPA) with the exact numerical evaluation of the WFs $\Psi_j(x)$ for the following sinusoidal double-well potential, shown in Fig.~\ref{figB1_tails_sin}(a):
\begin{equation}
\label{V(x)_pot_sin}
\cV(x)=
\begin{cases}
-V_0\sin |x| & \text{for} \quad |x|\le \pi\\
0 & \text{for} \quad |x|>\pi
\end{cases}\,,
\end{equation}
where $V_0>0$ is the potential depth. The potential in Eq.~\eqref{V(x)_pot_sin} can be seen as a sum of two single-well potentials, namely $\cV(x)=\cV_1(x)+\cV_2(x)$, where
\begin{equation}
\cV_1(x)=
\begin{cases}
-V_0\sin |x| & \text{for} \quad -\pi\le x\le 0\\
0 & \text{for} \quad x<-\pi \vee x>0
\end{cases}\,,
\label{V(x)_sinus}
\end{equation}
while $\cV_2(x)=\cV_1(-x)$.
\begin{figure*}[h]
\centering
{\includegraphics[width=17.2cm]{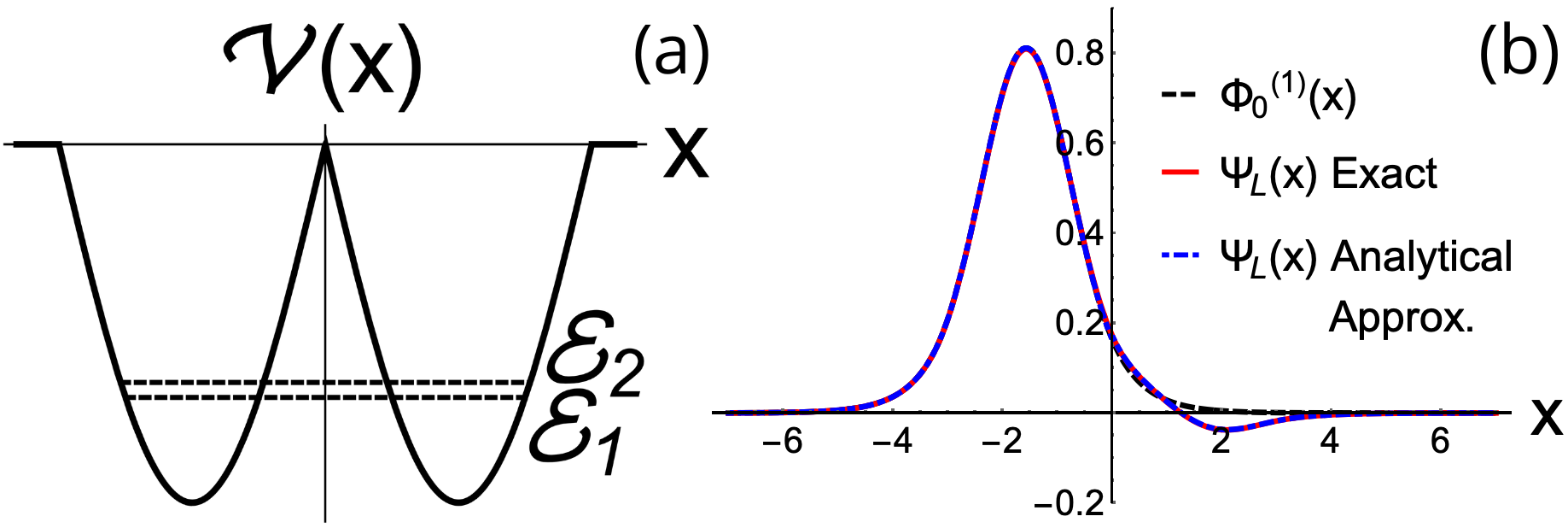}}
\caption{(a) Sinusoidal double-well potential with first two exact eigenenergies $\cE_{1,2}$ (black dashed lines). The bound states energies are: $\cE_1 \simeq -3.2$ and $\cE_2 \simeq -3.0$, respectively. (b) Comparison between the orbital $\Phi_0^{(1)}(x)$ (black dashed curve), the exact left-well WF $\Psi_L(x)$ (red solid curve) and the analytical approximated (neglecting $\cO(\Omega_0^2)$ terms) left-well WF (blue dot-dashed curve) for lattice depth $V_0=4.5$. The exact ground-state energy and tunneling coupling are $\bE_0 \simeq -3.1$ and $\bom_0 \simeq -0.094$, respectively, while the corresponding parameters given by the TPA are $E_0 \simeq -3.08$, $\Omega_0 \simeq -0.094$.}
\label{figB1_tails_sin}
\end{figure*}
To simplify the notation, in the following we use dimensionless units, i.e., $\hbar=2m=1$, unless otherwise specified. First, we consider the exact numerical solution for $E_\pm$ and the wave functions $\psi_\pm(E_\pm,x)$, given by Eqs.~\eqref{eigen_site} of the main text, where $\bPhi^{(1)}(E,x)$ is obtained from the numerical solution of the Schr\"odinger equation
\begin{equation}
\begin{aligned}
-&\bPhi^{(1)\prime\prime}(E,x)=\left(E-V_0\sin |x|\right)\bPhi^{(1)}(E,x) \quad {\rm for} \quad -\pi\le x\le 0\,,\\ 
&\bPhi^{(1)}(E,x)=e^{q (x+\pi)} \quad {\rm for}\quad 
x\le -\pi \,,
\end{aligned}
\end{equation}
where $q=\sqrt{-E}$. The eigenenergies are obtained from the matching conditions, namely $\bPhi^{(1)\prime}(E_+,0)=0$ for the symmetric state $\psi_+(E_+,x)$ and $\bPhi^{(1)}(E_-,0)=0$ for the antisymmetric state $\psi_-(E_-,x)$. Using Eqs.~\eqref{wan_approx} of the main text, we then obtain the exact WFs $\Psi_{L,R}(x)$. The eigenenergy levels $$E_\pm=\cE_{1,2}+\cO(\beta^2)$$ are denoted by dashed black lines in Fig.~\ref{figB1_tails_sin}(a), while the exact left-well WF $\Psi_L(x)$ is displayed in Fig.~\ref{figB1_tails_sin}(b) (red solid curve) for lattice depth $V_0=4.5$. The corresponding left-well orbital $\Phi_0^{(1)}(x)$, defined in Eqs.~\eqref{orbitals_N2} of the main text, is also shown (black dashed curve). We note that the WF almost corresponds to the respective orbital inside the respective well, while its tail in the other well differs from the orbital's one, as expected from our previous analysis. To compare these results with our approximation in Eqs.~\eqref{wan_approx} of the main text, we consider the tunneling Hamiltonian in Eq.~\eqref{tunn_ham_2part} of the main text, whose parameters $E_0$ and $\Omega_0$ are directly related to $E_\pm$ via Eq.~\eqref{Wannier_functions_A} and Eq.~\eqref{Wannier_functions_B} of the main text, see also Eqs.~\eqref{params_Ham} of the main text. Using $\Omega_0$ as energy shift in the orbital, we can evaluate the corresponding WF by using Eqs.~\eqref{wan_approx}, or equivalently Eqs.~\eqref{wan_approx_norm} of the main text. The resulting analytical approximated WF is represented in Fig.~\ref{figB1_tails_sin}(b) (blue dot-dashed curve). For this set of parameters, our analytical approximated approach produces a result which is very close to the exact one. We expect our approach to be the more accurate the more the barrier height ensures a good localization of the left and right states in the respective well.

\section{Dynamics of two distinguishable particles in a double-well potential}
\renewcommand{\theequation}{C.\arabic{equation}}
\setcounter{equation}{0}
\renewcommand{\thefigure}
{C.\arabic{figure}}
\setcounter{figure}{0}
\label{appendix_C}
In this section, we consider the effects of the nonstandard Hubbard terms on the dynamics of two distinguishable particles in the symmetric square double-well potential shown in Fig.~\ref{fig3_sdw} of the main text. As a figure of merit of the system's dynamics, we consider the oscillation frequency of the time-dependent probability $P_{LL}(t)$, defined as the occupation probability to find both particles in the left well at time $t$.
\begin{figure}[h] 
\centering
\includegraphics[width=17.2cm]{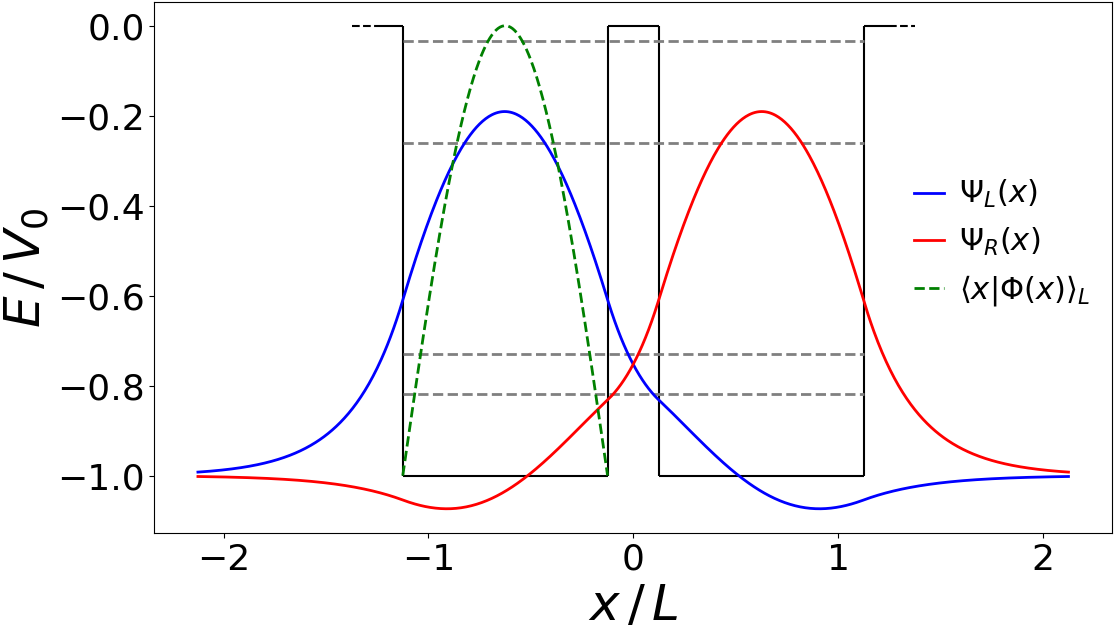}
\caption{Left (blue curve) and right (red curve) exact WFs for the symmetric square double-well system shown in Fig.~\ref{fig3_sdw} of the main text. Parameters are $L=2$, $b=0.5$ and $V_0=5$. The eigenvalues (gray dashed lines, from bottom to top) are: $E_0\simeq -4.089$, $E_1\simeq -3.644$, $E_2\simeq -1.308$, $E_3\simeq -0.169$. $\ket{\Phi(x)}_L$ (green dashed curve) is the symmetric ground state of the infinite left-well potential considered as initial state for the system's dynamics.}
\label{main_figure_well}
\end{figure}
As initial state for the dynamics, we consider both particles placed in the infinite left-well state, and we compute the time-dependent occupation probability $P_{LL}(t)$ for the state $\ket{LL} \equiv \hat{a}_L^{\dagger(1)}\hat{a}_L^{\dagger(2)}\ket{00}$. The system's parameters have been chosen in such a way to have four bound energies in the double-well system, as shown in Fig.~\ref{main_figure_well}.
We compute the dynamics of the two particles by employing the SB nonstandard model, where only one energy level per each well is considered, and the MB nonstandard model, where all the energy levels (inside the potential well) are taken into account.

\begin{figure*}[h]
\centering
{\includegraphics[width=17.2cm]{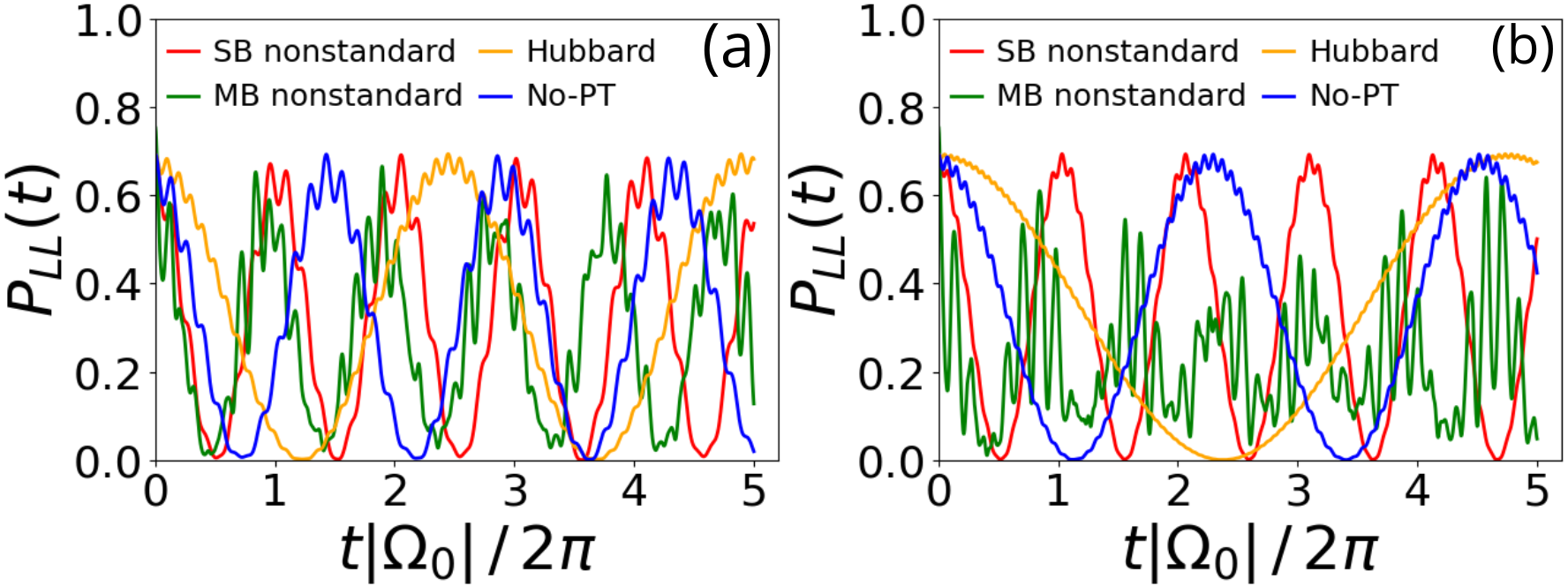}}
\caption{Time-evolution of the probability $P_{LL}(t)$ of finding at time $t$ two distinguishable particles together in the left well of the symmetric square double-well potential of Fig.~\ref{fig3_sdw} of the main text. The particles are interacting via $\delta$-shaped interaction potential of Eq.~\eqref{delta_potential} of the main text. The different models are: SB nonstandard model (red curve), MB nonstandard model (green curve), no-PT model (blue curve) and Hubbard model (orange curve). The double-well parameters are $L=2$, $b=0.5$ and $V_0=5$. The interaction strengths are: $U/|\Omega_0|\simeq 6$ (weakly interacting regime) in panel (a) and $U/|\Omega_0|\simeq 12$ (strongly interacting regime) in panel (b).}
\label{figC2_dynamics}
\end{figure*}

In Fig.~\ref{figC2_dynamics}, we show the probability $P_{LL}(t)$ to find the two particles in the left well, for two different repulsive interaction strengths, respectively in the weakly interacting regime in Fig.~\ref{figC2_dynamics}(a) and in the strongly interacting regime in in Fig.~\ref{figC2_dynamics}(b). Note that similar results were obtained for the case of attractive interaction strength in Fig.~\ref{fig7_dynamics} of the main text.

Specifically, we compute $P_{LL}(t)$ by using the standard Hubbard model, the SB nonstandard model and the MB nonstandard model, the last two including novel effects given by the exact treatment of the interaction. To stress the importance of the PT process, we also include the no-PT model, obtained discarding artificially the effect of the PT process itself on the system's dynamics from the SB nonstandard model.

Looking at Fig.~\ref{figC2_dynamics}, we note that the oscillation frequency of $P_{LL}(t)$ expected from the Hubbard model (yellow curve) is smaller than the one obtained from the SB and MB nonstandard models (respectively red and green curves), for both weakly and strongly repulsive interacting regimes. This highlights the role of the exact treatment of interaction in the system's dynamics, justifying the inclusion of the nonstandard terms in the Hamiltonian of the system. Specifically, we note that for weakly interacting regime, the frequency is closer to the SB nonstandard model one, being the nonstandard term less relevant in the dynamics, compared with the strongly interacting regime. Finally, we appreciate the difference between SB and MB nonstandard models, given by the different number of energy levels included in the dynamics, especially in the strongly interacting regime (the excitation probability for the higher levels is increased).

\section{Dynamics of two bosons in a sinusoidal double-well potential: numerical analysis and experimental comparisons}
\renewcommand{\theequation}{D.\arabic{equation}}
\setcounter{equation}{0}
\renewcommand{\thefigure}
{D.\arabic{figure}}
\setcounter{figure}{0}
\label{appendix_D}
 
In this section, we analyze the dynamics of two bosons in a sinusoidal double-well potential in the parameter range between weakly and strongly interacting regimes, see Ref.~\cite{bloch:direct_observation}. Using our numerical simulations, and exploiting the versatility of our numerical method, which allows us to solve the dynamics for different potential shapes, we investigate regions of parameters relevant to the experimental results reported in Ref.~\cite{bloch:direct_observation}. Unless stated otherwise, all energies are expressed in units of the recoil energy $$E_R=\frac{h^2}{2m\lambda^2}\,,$$
where $\lambda=765 \ nm$ represents the periodicity of the lattice, and $m$ is the mass of the atoms (see Sec.~\ref{sec:COMP_EXP} in the main text).

Restricting our considerations to a non-tilted double-well potential, the bosonic atoms are described by the standard Bose-Hubbard Hamiltonian, see Ref.~\cite{bloch:direct_observation}:
\begin{equation}
\hat{H}_{BH}=-J\left(\hat{b}_L^{\dagger}\hat{b}_R+\hat{b}_R^{\dagger}\hat{b}_L\right)+\frac{U}{2}\left[\hat{n}_L\left(\hat{n}_L-1\right)+\hat{n}_R\left(\hat{n}_R-1\right)\right]\,,
\end{equation}
where $J\equiv \Omega_0$ is the single-particle tunneling rate, $\hat{b}_{L,R}^{(\dagger)}$ are the annihilation (creation) operators for a bosonic particle in the ground state of the left and right wells, and $U$ represents the interaction energy between two particles within the same well. The experimental sinusoidal double-well potential in Ref.~\cite{bloch:direct_observation} is achieved by superimposing two optical lattices with different periodic potentials: a long-lattice potential with amplitude $A_{long}$ and wavelength 765.0 nm, and a short-lattice potential with amplitude $B_{short}$ and wavelength 382.5 nm. The dynamics is initiated with both bosons placed in the left well, and a sudden decrease in the potential barrier depth triggers their motion. The system's dynamics is characterized by the average position
\begin{equation}
\braket{x(t)}=\frac{\braket{\hat{n}_R(t)}-\braket{\hat{n}_L(t)}}{\braket{\hat{n}_R(t)}+\braket{\hat{n}_L(t)}}\,,
\end{equation}
where $\hat{n}_{L,R}$ denotes the number of bosons in the left (L) and right (R) wells, respectively. We consider regions of parameters near to the experimental results reported in Ref.~\cite{bloch:direct_observation}, namely both weakly ($U/J=0.67$) and strongly ($U/J=5$) interacting regimes. Our numerical model extends the standard Bose-Hubbard Hamiltonian to a single-band (SB) nonstandard Hubbard Hamiltonian, described as:
\begin{equation}
\begin{aligned}
\hat{H}=&-J\left(\hat{b}_L^{\dagger}\hat{b}_R+\hat{b}_R^{\dagger}\hat{b}_L\right)+\frac{U}{2}\left[\hat{n}_L\left(\hat{n}_L-1\right)+\hat{n}_R\left(\hat{n}_R-1\right)\right]+\bU\left(\hat{n}_L\hat{n}_R\right)\\
&+\Omega_1\left[\hat{b}^{\dagger}_L\left(\hat{n}_L+\hat{n}_R\right)\hat{b}_R+\hat{b}^{\dagger}_R\left(\hat{n}_R+\hat{n}_L\right)\hat{b}_L\right]\\
&+\Omega_2\left(\hat{b}^{\dagger 2}_L\hat{b}^2_R+\hat{b}^{\dagger 2}_R\hat{b}^2_L\right)\,,
\end{aligned}
\end{equation}
where $\bU$ represents the nearest-neighbor interaction, $\Omega_1$ denotes the density-induced tunneling (DT) term, and $\Omega_2$ represents the pair tunneling (PT) term.
\begin{figure}[h]
\centering
\includegraphics[width=17.2cm]{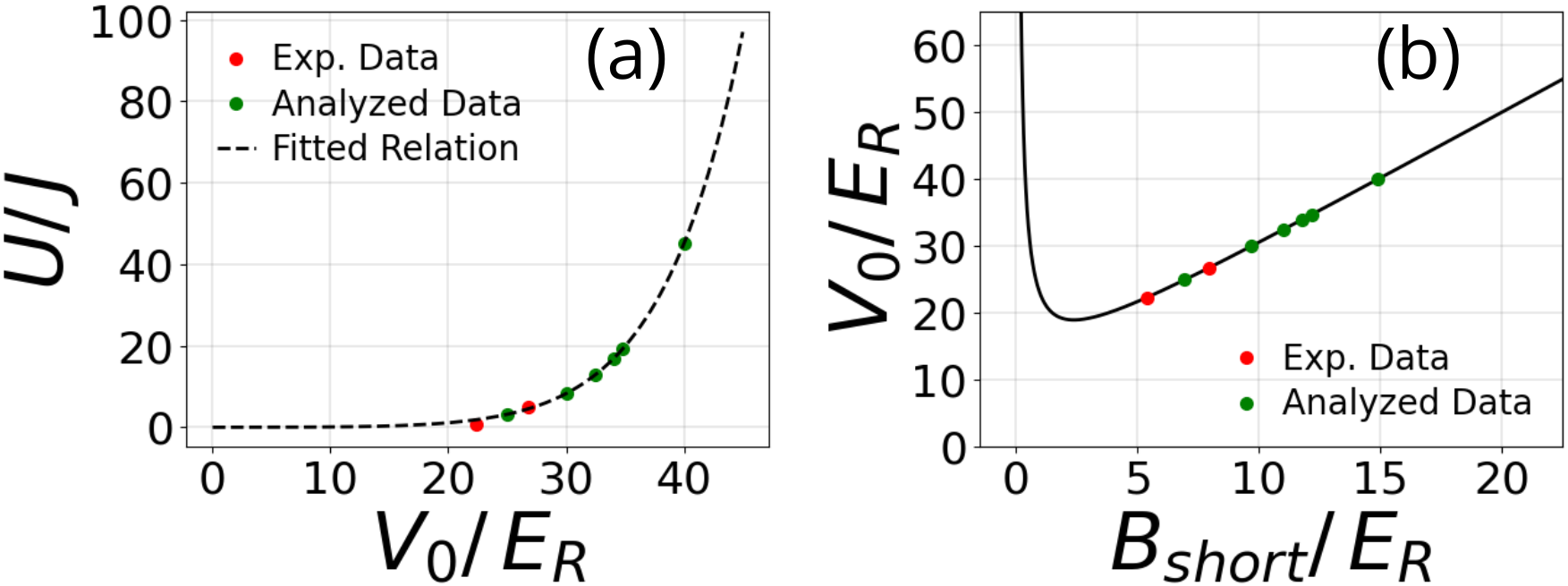}
\caption{Panel (a): exponential relation between the interaction-to-tunneling ratio $U/J$ and the lattice depth $V_0/E_R$ as described by Eq.~\eqref{eq_U_J}. Red dots: parameter regimes for weakly ($U/J=0.67$) and strongly ($U/J=5$) interacting bosons, as taken from Ref.~\cite{bloch:direct_observation}. Green dots: data utilized in our numerical simulations ($U/J= 3.19, 8.27, 12.94, 16.78, 19.24, 45.04$) and ($V_0/E_R= 25, 30, 32.5, 34, 34.8, 40$). Panel (b): relation between the lattice depth $V_0$ and the short-lattice amplitude $B_{short}$. The short-lattice amplitude $B_{short}$ is varied to mimic experimental conditions using two superimposed optical periodic potentials.}
\label{fig_D1}
\end{figure}

To simulate experimental conditions, we vary the ratio $U/J$ as a function of the lattice depth $V_0$. From Ref.~\cite{manybody_ultracold}, we know that the ratio $U/J$ increases exponentially with the lattice depth $V_0$, as follows:
\begin{equation}
\label{eq_U_J}
\frac{U}{J}\sim \frac{a}{d}\exp \left(\sqrt{4V_0/E_R}\right)\,,
\end{equation}
where $a$ is the scattering length and $d=\lambda/2$ is the lattice spacing (see Sec.~\ref{sec:COMP_EXP} in the main text).
\begin{figure}[h]
\centering
\includegraphics[width=17.2cm]{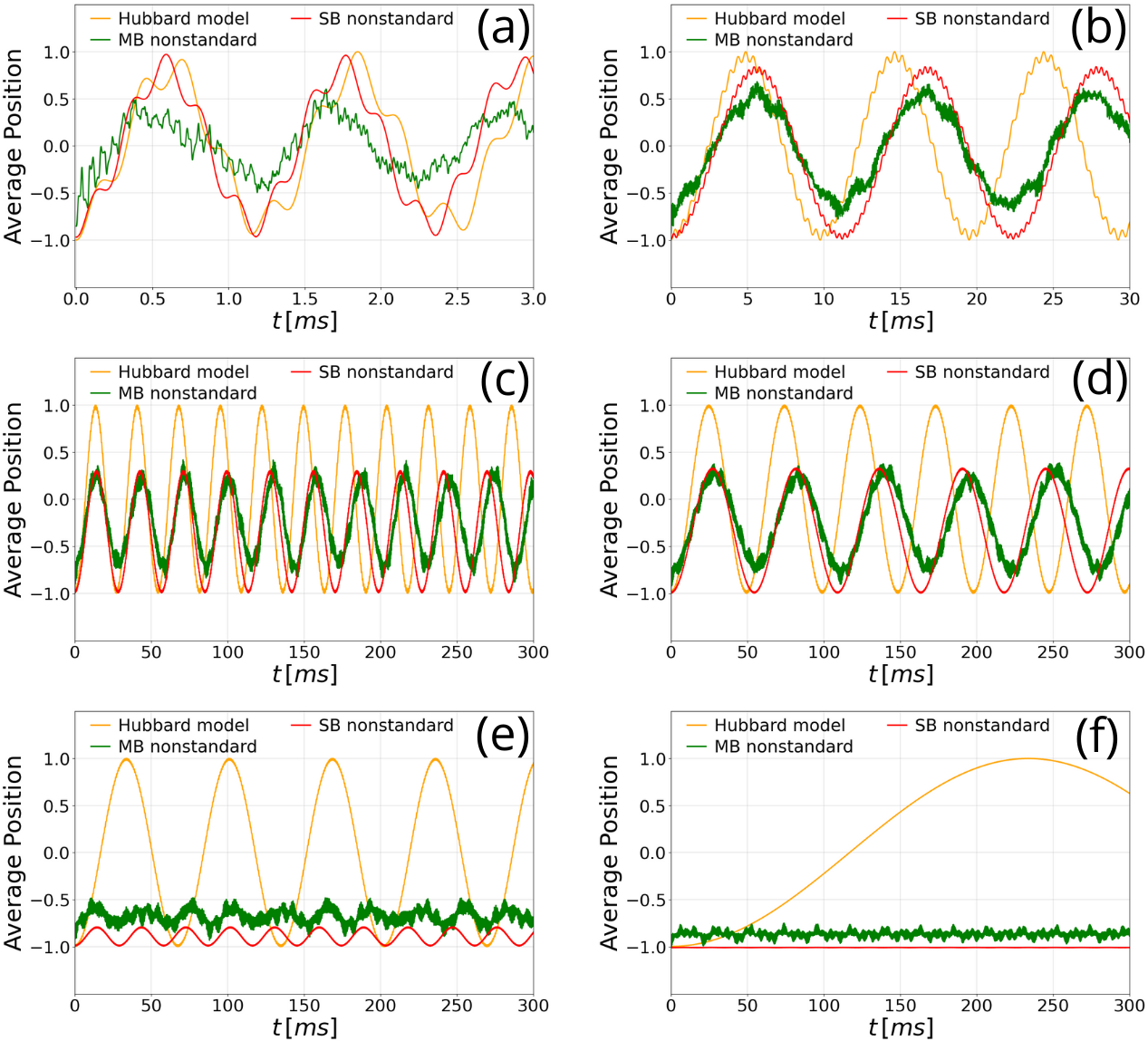}
\caption{Average position of two interacting bosons as a function of time. The dynamics of two bosons initially placed in the left well are simulated for different interaction regimes, with varying $U/J$ ratios (from Fig.~\ref{fig_D1}). The plot compares the average position predicted by the standard Hubbard model (orange curve), the SB nonstandard model (red curve), and the MB nonstandard model (green curve). Panel (a): $|\Omega_1/\Omega_0|=0.13$, $|\Omega_2/\Omega_0|=0.04$, $U/\Delta E=0.15$. Panel (b): $|\Omega_1/\Omega_0|=0.12$, $|\Omega_2/\Omega_0|=6\cdot 10^{-3}$, $U/\Delta E=0.08$. Panel (c): $|\Omega_1/\Omega_0|=0.17$, $|\Omega_2/\Omega_0|=3\cdot 10^{-3}$, $U/\Delta E=0.06$. Panel (d): $|\Omega_1/\Omega_0|=0.08$, $|\Omega_2/\Omega_0|=2\cdot 10^{-3}$, $U/\Delta E=0.05$. Panel (e): $|\Omega_1/\Omega_0|=0.08$, $|\Omega_2/\Omega_0|=2\cdot 10^{-3}$, $U/\Delta E=0.04$. Panel (f): $|\Omega_1/\Omega_0|=3.23$, $|\Omega_2/\Omega_0|=0.12$, $U/\Delta E=0.03$.}
\label{fig_D2}
\end{figure}
Fig.~\ref{fig_D1}(a) illustrates the relation in Eq.~\eqref{eq_U_J}, highlighting both the weakly and strongly interacting regimes from Ref.~\cite{bloch:direct_observation} (red dots) and our analyzed data (green dots). Fig.~\ref{fig_D1}(b) shows the relation between lattice depth $V_0$ and the short-lattice amplitude $B_{short}$.

The average position of two interacting bosons, initially placed in the infinite left-well state, for various values of $U/J$ and corresponding lattice depths $V_0$, spanning regions near to weakly ($U/J=0.67$) and strongly ($U/J=5$) interacting regimes, is shown in Fig.~\ref{fig_D2}. We compare the standard Hubbard model (orange curve) with the SB nonstandard model (red curve). For further comparison, we also include the multiband (MB) nonstandard model (green curve, obtained numerically) to highlight the region where the SB approximation holds. As one can see, the nonstandard model predicts significantly different average positions compared with the standard Hubbard model, both in amplitude and frequency, underscoring the importance of nonstandard terms in capturing the system's dynamics. Moreover, in all panels of Fig.~\ref{fig_D2}, it is clear that the SB approximation is quite accurate for the interaction values considered. Nevertheless, it is important to observe that the SB nonstandard model closely approximates the MB nonstandard model when $U \ll \Delta E$ (i.e., when the interaction energy $U$ does not exceed $\Delta E$, ensuring that the energy levels are not widely spaced).
\begin{figure}[t]
\centering
\includegraphics[width=17.2cm]{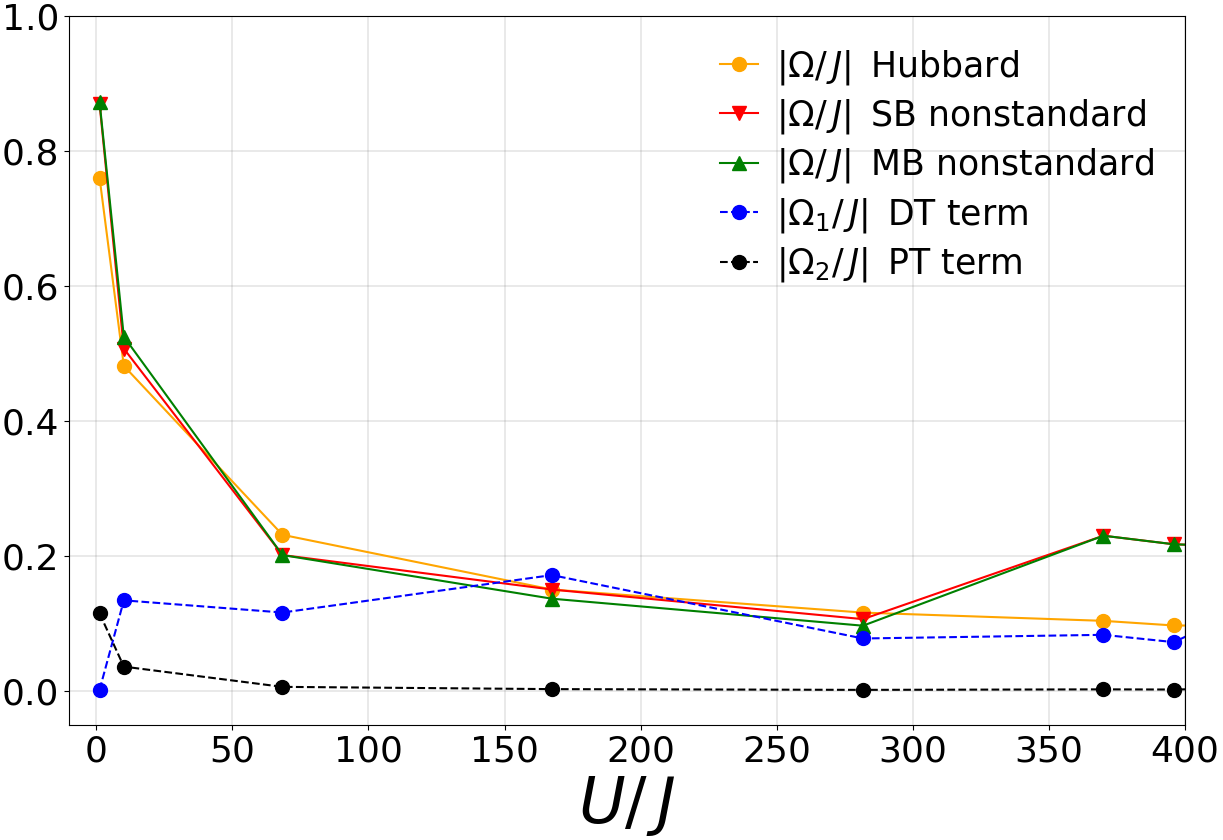}
\caption{Frequency spectrum of the dynamics of two bosons for different models, as a function of the interaction-to-hopping parameter $U/J$. The plot shows the frequencies of oscillation for the system as predicted by the standard Hubbard model, the SB nonstandard model, and the MB nonstandard model. The density-induced tunneling (DT) term $\Omega_1$ and the pair tunneling (PT) term $\Omega_2$ are also displayed.}
\label{fig_D3}
\end{figure}

A sharp transition from low to high $U/J$ values is evident in Fig.~\ref{fig_D2}. Specifically, at low $U/J$ values, the bosons tunnel between wells (metallic phase), while for large enough $U/J$, they remain localized in the initial well (insulating phase). This transition occurs in both the standard Bose-Hubbard model and the nonstandard models, but at considerably larger $U/J$ values. As $U/J$ increases, both the oscillation frequency and the amplitude of oscillation of the average position decrease (see the yellow curve for the Hubbard model), highlighting the emergence of an insulating phase.

In the SB and MB nonstandard models, the metal-insulator transition still occurs (see the red and green curves, respectively), but at a lower \lq\lq critical\rq\rq $U/J$ value compared with the standard Hubbard model. This shift is primarly due to the effect of DT on the single-particle tunneling. An effective single-particle tunneling amplitude, $\Omega_{eff}=\Omega_0+\Omega_1$, arises, which is interaction-dependent, as detailed in Refs.~\cite{dutta:non_standard, jurgensen:observation_density_induced_tunnelling}. For repulsive interaction, $\Omega_0$ is reduced by $\Omega_1$, resulting in a lower $\Omega_{eff}$ and, consequently, a shifted phase transition with respect to $U/J$. To confirm such a view, Fig.~\ref{fig_D3} shows the dynamical frequencies predicted by the standard Hubbard model and both the SB and MB nonstandard models, along with the DT term $\Omega_1$ and PT term $\Omega_2$. Notably, the latter is largely irrelevant in the transition to the localized phase.

In conclusion, under the given experimental conditions, it is challenging to find interesting experimental parameter regions where the PT term is significant. This is because increasing $U$ also necessitates increasing $V_0$, at variance with the theoretical case discussed. Despite this limitation, the presence of DT term requires a significant modification of the critical interaction strength at which the insulating phase appears. This suggests that the metal-insulator transition could be highly sensitive to the additional nonstandard terms considered in this work. On the other hand, alternative experimental setups may be needed to explore the relevance of the PT term further.

\end{document}